\documentclass[fleqn,usenatbib,useAMS]{mnras}

\usepackage{newtxtext,newtxmath}

\usepackage[T1]{fontenc}
\usepackage{ae,aecompl}

\usepackage{color}
\usepackage{xspace}
\usepackage{graphicx}	
\usepackage{amsmath}	
\usepackage{amssymb}	
\usepackage{siunitx}
\usepackage{multirow}
\usepackage{array,booktabs}
\usepackage[flushleft]{threeparttable}
\usepackage{ulem}
\usepackage{breakcites}
\usepackage[disable,colorinlistoftodos,textsize=7]{todonotes}
\newcommand{\gray}[1]{\textcolor{gray}{#1}}
\newcommand{\ie}{{\it i.e.}}
\newcommand{\eg}{{\it e.g.}}
\newcommand{\pcc}{~{\rm cm}^{-3} }	
\newcommand{\tff}{t_{\rm ff}} 
\newcommand{\md}{\mathrm{d}} 
\newcommand{\HII}{\mbox{H\,{\sc ii}} } 
\newcommand{\HeII}{\mbox{He\,{\sc ii}} } 

\newcommand{\fesc}{\ensuremath{\langle f_{\rm esc}^{\scriptscriptstyle \rm MC}\rangle}\xspace} 
\newcommand{\fescg}{\ensuremath{\langle f_{\rm esc}^{\rm gal}\rangle}\xspace} 
\newcommand{\fesci}{\ensuremath{f_{\rm esc}^{\scriptscriptstyle \rm ISM}}}
\newcommand{\msun}{\ensuremath{\,{\rm M}_{\odot}}} 

\newcommand{\ltsima}{\mbox{$\; \buildrel < \over \sim \;$}}
\def\simlt{\lower.5ex\hbox{\ltsima}}
\def\gtsima{$\; \buildrel > \over \sim \;$}
\def\simgt{\lower.5ex\hbox{\gtsima}}
\def\hide#1{}

\begingroup
\catcode`\_=\active
\gdef_#1{\ensuremath{\sb{\mathrm{#1}}}}
\endgroup
\mathcode`\_=\string"8000
\catcode`\_=12

\title[Star Clusters Across Cosmic Time]
{Simulating Star Clusters Across Cosmic Time: II. Escape Fraction of Ionizing Photons from Molecular Clouds}
\author[C.-C. He, M. Ricotti, S. Geen]{
Chong-Chong He,$^{1}$\thanks{E-mail: chongchong@astro.umd.edu}
Massimo Ricotti,$^{1}$\thanks{E-mail: ricotti@umd.edu}
and Sam Geen$^{2}$\\
$^{1}$Department of Astronomy, University of Maryland, College Park, MD, 20742, US\\
$^{2}$Universit\"at Heidelberg, Zentrum f\"ur Astronomie, Institut f\"ur Theoretische Astrophysik, Albert-Ueberle-Str. 2, 69120 Heidelberg, Germany
}

\date{Accepted XXX. Received YYY; in original form ZZZ}

\pubyear{2018}

\begin{document}

\label{firstpage}
\pagerange{\pageref{firstpage}--\pageref{lastpage}}
\maketitle

\begin{abstract}
We calculate the hydrogen and helium-ionizing radiation escaping star forming molecular clouds, as a function of the star cluster mass and compactness, using a set of high-resolution radiation-magneto-hydrodynamic simulations of star formation in self-gravitating, turbulent molecular clouds. In these simulations, presented in He, Ricotti and Geen (2019), the formation of individual massive stars is well resolved, and their UV radiation feedback and lifetime on the main sequence are modelled self-consistently.
We find that the escape fraction of ionizing radiation from molecular clouds, $\langle f_{\rm esc}^{\scriptscriptstyle \rm MC}\rangle$, decreases with increasing mass of the star cluster and with decreasing compactness. Molecular clouds with densities typically found in the local Universe have negligible $\langle f_{\rm esc}^{\scriptscriptstyle \rm MC}\rangle$, ranging between $0.5\%$ to $5\%$. Ten times denser molecular clouds have $\langle f_{\rm esc}^{\scriptscriptstyle \rm MC}\rangle \approx 10\%-20\%$, while $100\times$ denser clouds, which produce globular cluster progenitors, have $\langle f_{\rm esc}^{\scriptscriptstyle \rm MC}\rangle \approx 20\%-60\%$. 
We find that $\langle f_{\rm esc}^{\scriptscriptstyle \rm MC}\rangle$ increases with decreasing gas metallicity, even when ignoring dust extinction, due to stronger radiation feedback. However, the total number of escaping ionizing photons decreases with decreasing metallicity because the star formation efficiency is reduced.
We conclude that the sources of reionization at $z>6$ must have been very compact star clusters forming in molecular clouds about $100\times$ denser than in today's Universe, which lead to a significant production of old globular clusters progenitors.
\end{abstract}

\begin{keywords}
keyword1 -- keyword2 -- keyword3
\end{keywords}


\pagebreak

\section{Introduction}

A large observational effort is underway to understand the epoch of reionization, both by observing the high-redshift sources of radiation with HST and JWST \citep{Ellis2013, Sharma2016, Oesch2016} and detecting the 21cm signal from neutral hydrogen in the intergalactic medium (IGM) \cite[\eg,][]{Bowman:2018}.
Numerical simulations of galaxy formation are becoming increasingly realistic, but the question of which are the sources that propelled reionization is largely unanswered. To answer this question it is necessary to know the mean value of the escape fraction of ionizing radiation, \fescg, from dwarf and normal galaxies into the IGM at redshift $z>6$. This quantity is arguably the most uncertain parameter in models of reionization. 
It is difficult to measure, and for the cases in which it has been measured in galaxies at $z \approx 1$, upper limits of $f_{esc} \approx 2$ per cent has been typically found \citep[][\eg,]{Bridge2010}.
Using staking techniques in Lyman-break galaxies at $z\sim 3$ some authors claimed higher values of $f_{esc}$ at 5--7 per cent \citep{Vanzella2012, Nestor2013}. However, according to simulations of reionization a mean value of \fescg$\simgt 10-20\%$ is required to reionize the IGM by $z\sim 6.2$ \citep{Ouchi:2009,Robertson:2015,Khaire:2016}. This value is too large with respect to what observed in local galaxies, unless at high-redshift the value of \fescg is significantly larger than in the local Universe.

Recently, a handful of galaxies at high redshifts have been confirmed to have large Lyman continuum (LyC) escape fractions. 
{\it Ion2} and {\it Q1549-C25} are the only two $z \sim 3$ galaxies with a direct spectroscopic detection of uncontaminated LyC emission \citep{Vanzella:2016, Shapley:2016}. Escape fractions of $\gtrsim 50\%$ is inferred for both of them. \cite{Vanzella:2018} reported the highest redshift individually-confirmed LyC-leaky galaxy, {\it Ion3}, at $z = 4$.
As a proxy for high-z galaxies, \cite{Izotov:2018} selected local compact star-forming galaxies in the  redshifts range $z = 0.2993 - 0.4317$, using the Cosmic Origins Spectrograph on HST.  They found LyC emission with $f_{esc}$ in a range of 2-72 per cent. We should note that \fescg in models of reionization is the averaged value over all star forming galaxies, but also a time-average of $f_{esc}(t)$ over the duration of the starburst.

A number of attempts have been made to predict the escape fraction of hydrogen LyC photons from galaxies using analytic models and simulations of galaxy formation \citep{Ricotti:2000, Gnedin2008, Wise:2009, Razoumov2010, Yajima:2011, Wise:2014, Ma2015, Xu:2016},
but because of the complexity of the problem and the uncertainty about the properties of the sources of reionization, the results are inconclusive. In addition, any realistic theoretical estimate of \fescg must take into account the escape fraction of ionizing radiation from the molecular clouds in which the stars are born, \fesc, a sub-grid parameter in galaxy-scale and in cosmological-scale simulations. Typically \fesc is set to unity in cosmological simulations of reionization, which could dramatically overpredict \fescg \citep[\eg,][]{Ma2015}. More recent simulations which do not make a priori assumptions about subgrid escape fractions \citep[\eg,][]{Rosdahl:2018} remain very sensitive to small-scale effects. In addition, they require that outflows from star-forming regions clear channels in the galaxies while ionising radiation is still being emitted in large enough quantities, for example by invoking binary stellar evolution models.

A small body of work exists that estimates \fesc in star-forming molecular clouds \citep{Dale:2014,Howard:2017,Howard:2018,Kimm:2019}, although systematic studies remain limited in number.
\cite{Dale:2014} finds that \fesc$\propto 1/L_{\rm cl}$, or that the escaping ionizing radiation rate from star clusters of different masses is roughly constant at a few $\times 10^{49} \ {\rm s}^{-1}$. However, in this work the calculation of \fesc assumes that all the radiation is emitted from a point source located at the center of the cloud. Also, in this work the clouds have the same initial density, similar to today's molecular clouds associated with young star forming regions.
\cite{Howard:2018} find the overall escape fraction is not a monotonic function of the cloud mass, $m_{\rm gas}$, varying from $31\%$ for $m_{\rm gas}=10^4\msun$, to $100\%$ for $m_{\rm gas}=10^5\msun$, and $9\%$ for from $m_{\rm gas}=10^6\msun$. They also use a rather crude estimation of \fesc in their simulations by assuming that all the radiation is emitted from a point source located at the center of the star cluster. Observationally, escape fractions from molecular clouds remain uncertain. \cite{Doran:2013} find an escape fraction of ionising photons of 6\% from 30 Doradus in the Large Magellanic Cloud, but their error bars give a maximum possible escape fraction of 71\%.

In this paper, the second of a series, we estimate \fesc using a large set of realistic simulations of star cluster formation in molecular clouds. These are radiation-magneto-hydrodynamic simulations of star formation in self-gravitating, turbulent molecular clouds, presented in \cite*{He2019} (hereafter, Paper~I). We model self-consistently the formation of individual massive stars, including their UV radiation feedback and their lifetime. We consider a grid of simulations varying the molecular cloud masses between $m_{gas}=10^3$~M$_\odot$ to $3 \times 10^5$~M$_\odot$, and resolving scales between 200~AU to 2000~AU. We also varied the compactness of the molecular clouds, with mean gas number densities typical of those observed in the local Universe ($\overline n_{gas} \sim \num{1.8e2}$~cm$^{-3}$) and denser molecular clouds ($\overline n_{gas}\sim \num{1.8e3}$~cm$^{-3}$ and $\num{1.8e4}$~cm$^{-3}$) expected to exist, according to cosmological simulations \citep{Ricotti:2016}, in high-redshift galaxies. We also partially explored the effects of varying the gas metallicity.

Previous works have suggested that the progenitors of today's old globular clusters, and more generally compact star cluster formation, may have been the dominant mode of star formation before the epoch of reionization, and that GC progenitors may have dominated the reionization process \citep{Ricotti:2002, KatzR:2013, KatzR:2014, Schaerer:2011, Boylan-Kolchin:2018}.
\cite{Ricotti:2002} have shown that if a non-negligible fraction of today's GCs formed at $z>6$ and had $\fesc \sim 1$, they would be a dominant source of ionizing radiation during reionization. \cite{KatzR:2013} presented arguments in support of significant fraction of today's old GCs forming before the epoch of reionization. However, although it seems intuitive, it has not been shown that \fesc from proto-GCs forming in compact molecular clouds is higher than \fesc in more diffuse clouds. Answering this question, and quantifying the contribution of compact star clusters to reionization is a strong motivation for this work.

In a scenario in which the progenitors of today's GCs dominate the reionization process, we expect a short effective duty cycle in the rest-frame UV bands, leading to a large fraction of halos of any given mass being nearly dark in between short-lived bursts of star formation. In addition, large volumes of the universe would be only partially ionized inside relic \HII regions produced by bursting star formation. \cite{Hartley:2016} have shown that the number of recombinations and therefore the number of ionizing photons necessary to reionize the IGM by $z=6.2$ is lower in this class of models with short bursts of star formation with respect to models in which star formation is continuous (producing fully ionized \HII bubbles). In summary, for the reasons discussed above, compact star clusters are a very favorable candidate to propel reionization: i) deep field surveys of sources at $z>6$ suggest that the sources of reionization are a numerous but faint population. Compact star clusters would fit this requirement, also due to their their short duty cycle. ii) The value of \fesc necessary for reionization is reduced if star formation is bursty. iii) We naively expect that compact star clusters have higher star formation efficiency (SFE) and \fesc than less compact star clusters. This last point is the focus of this paper.

This paper is organised as follows. In Section~\ref{sec:method} we present the simulations and the analysis methods. Section~\ref{sec:analyse} presents all the results from the numerical simulations regarding \fesc, while in Section~\ref{sec:disc} we discuss the physical interpretation of the results and their analytical modelling. We also discuss the implications for reionization assuming a simple power-law distribution of the cluster masses, similar to what is observed in the local universe. A summary of the results and conclusions are in Section~\ref{sec:summary}.

\section{Numerical Simulations and Methods}
\label{sec:method}

\subsection{Simulations}
\label{sec:sims}

\begin{table*}
  \caption{A table of parameters in all simulations.}
  \label{tab:1}
  \begin{threeparttable}
    \centering \def\arraystretch{1.1}
    \begin{tabular}{cccccccSS}
      \toprule
      Compactness
      & Cloud Name
      & $m_{gas}$ (\msun) \tnote{a}
      & $\overline n_{gas}$ (cm$^{-3}$) \tnote{b}
      & {$\Sigma$ ($\msun~{\rm pc}^{-2}$) \tnote{c}}
      & $Z~(Z_\odot$) \tnote{d}
      & Photon bins
      & {$t_{ff}$ (Myr) \tnote{e}}
      & {$t_{cr}$ (Myr) \tnote{f}} \\ 
      \midrule
      Fiducial & XS-F     & \num{3.2e+03} & \num{1.8e+02} & 41  & 1 & H, He, He$^+$ & 4.4 & 0.5 \\
      Fiducial & S-F      & \num{1.0e+04} & \num{1.8e+02} & 61  & 1 & H, He, He$^+$ & 4.4 & 0.7 \\
      Fiducial & M-F      & \num{3.2e+04} & \num{1.8e+02} & 89  & 1 & H, He, He$^+$ & 4.4 & 1.1 \\
      Fiducial & L-F      & \num{1.0e+05} & \num{1.8e+02} & 131 & 1 & H, He, He$^+$ & 4.4 & 1.5 \\
      Fiducial & XL-F     & \num{3.2e+05} & \num{1.8e+02} & 193 & 1 & H, He, He$^+$ & 4.4 & 2.3 \\
      \hline
      Compact & XS-C     & \num{3.2e+03} & \num{1.8e+03} & 193 & 1 & H, He, He$^+$ & 1.4 & 0.23 \\
      Compact & S-C      & \num{1.0e+04} & \num{1.8e+03} & 283 & 1 & H    & 1.4 & 0.33 \\
      Compact & M-C      & \num{3.2e+04} & \num{1.8e+03} & 415 & 1 & H    & 1.4 & 0.5 \\
      Compact & L-C      & \num{1.0e+05} & \num{1.8e+03} & 609 & 1 & H    & 1.4 & 0.7 \\
      Compact & L-C-lm   & \num{1.0e+05} & \num{1.8e+03} & 609 & 1/10 & H & 1.4 & 0.7 \\
      Compact & L-C-xlm  & \num{1.0e+05} & \num{1.8e+03} & 609 & 1/40 & H & 1.4 & 0.7 \\
      \hline
      Very Compact & XXS-VC   & \num{1.0e+03} & \num{1.8e+04} & 609  & 1 & H, He, He$^+$ & 0.44 & 0.07 \\
      Very Compact & XS-VC    & \num{3.2e+03} & \num{1.8e+04} & 894  & 1 & H & 0.44 & 0.1 \\
      Very Compact & S-VC     & \num{1.0e+04} & \num{1.8e+04} & 1312 & 1 & H & 0.44 & 0.15 \\
      Very Compact & M-VC     & \num{3.2e+04} & \num{1.8e+04} & 1925 & 1 & H & 0.44 & 0.23 \\
      Very Compact & L-VC     & \num{1.0e+05} & \num{1.8e+04} & 2827 & 1 & H & 0.44 & 0.33 \\
      \bottomrule
    \end{tabular}
    \begin{tablenotes}
        \item[]
  (a) Initial cloud mass, excluding the envelope.
  (b) Mean number density of the cloud, excluding the envelope. The core density is $\sim 5$ times higher.
  (c) The mean surface density in a square of the size of the cloud radius.
  (d) Metallicity of the gas used in the cooling function, $Z$~=~[Fe/H].
  (e) The global free-fall time of the cloud ($t_{ff} \equiv 3\sqrt{\frac{3\pi}{32G\rho_{\rm c}}} \approx 1.3 \sqrt{\frac{3\pi}{32G\overline{\rho}}})$.
  (f) Sound crossing time $r_{\rm gas}/c_s$ with $c_s=10$~km/s.
    \end{tablenotes}
  \end{threeparttable}
\end{table*}

The results presented in this paper are based on a grid of 14 simulations of star formation in molecular clouds with a range of initial gas densities and masses, and 2 simulations varying the initial gas metallicity. For details about the simulations and main results regarding the IMF, star formation efficiency and star formation rate, we refer to Paper~I. Here, for the sake of completeness, we briefly describe the main characteristic of the code we used, and the simulations set up.

We run the simulations using an Adaptive Mesh Refinement radiative magneto-hydrodynamical code {\sc ramses} \citep{Teyssier:2002, Bleuler:2014}. 
Radiative transfer is implemented using a first-order moment method described in \cite{Rosdahl:2013}. The ionising photons interact with neutral gas and we track the ionization state and cooling/heating processes of hydrogen and helium. We include magnetic fields in the initial conditions. We do not track the chemistry of molecular species.

We simulate a set of isolated and turbulent molecular clouds that collapse due to their own gravity. The clouds have initially a spherically symmetric structure with density profile of a non-singular isothermal sphere with core density $\rho_{\rm c}$.  The initial density profile is perturbed with a Kolmogorov turbulent velocity field with an amplitude such that the cloud is approximately in virial equilibrium. A summary of the parameters of the simulations is presented in Table~\ref{tab:1}.

Proto-stellar cores collapsing below the resolution limit of the simulations produce sink particles. These sinks represent molecular cloud cores in which we empirically assume that fragmentation leads to formation of a single star with a mass roughly $40\%$ of the mass of the sink particle, and the remaining $60\%$ of the mass fragments into smaller mass stars. With this prescription we reproduce the slope and normalization of the IMF at the high-mass end. Stars emit hydrogen and helium ionising photons according to their mass using \cite{Vacca:1996} emission rates with a slight modification. We extend the high-mass-end power-law slope down to about $1 \,\msun$, therefore increasing the feedback of stars with masses between 1 and $30\,\msun$\footnote{This modification was an unintended result of a coding error, but further investigations have shown that it is important in producing the correct slope of the IMF.}.
The gas is ionized and heated by massive stars, producing over-pressurised bubbles that blow out the gas they encounter. In our simulations low mass stars and proto-stellar cores do not produce any feedback. We do not include mechanical feedback from supernova (SN) explosions and from stellar winds and we also neglect the effect of radiation pressure from infrared radiation. However, with the exception of a sub-set of simulations representing today's molecular clouds (the two most massive clouds in lowest density set), all the simulations stop forming stars before the explosion of the first SN. Therefore, neglecting SN feedback is  well justified in these cases.

\begin{figure*}
  \centering
  \includegraphics[width=\textwidth]{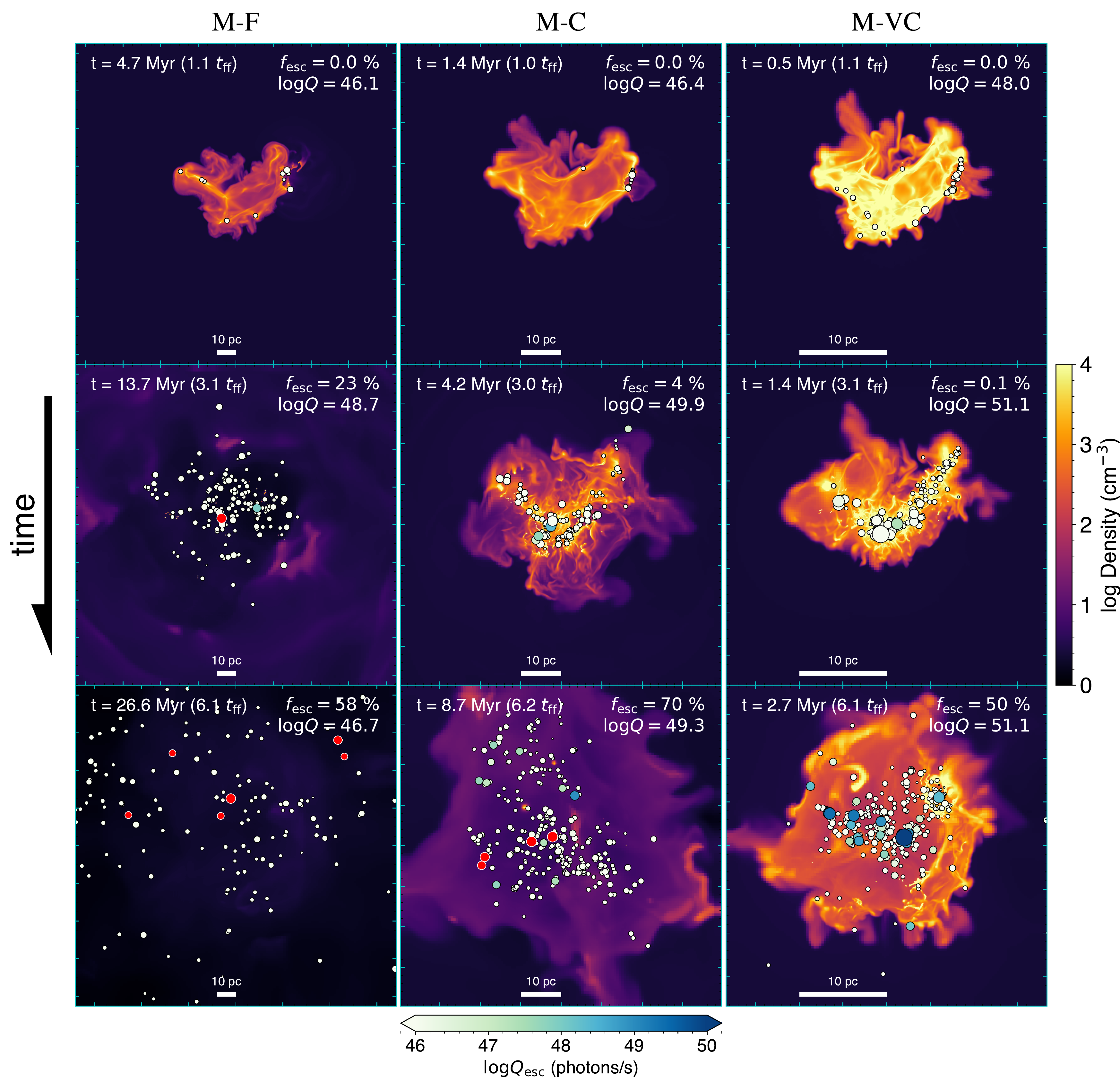}
  \caption{
  Time sequence plot of line-of-sight projections of density-weighted gas density from the Medium mass-Fiducial (M-F), Compact (M-C), and Very Compact (M-VC) clouds. These clouds have initial mass of $\num{3e4}\msun$ and initial mean density of $\num{2e2}$, $\num{2e3}$, and $\num{2e4}$ cm$^{-3}$.
  Sink particles are plotted as filled circles on top of the density map. These circles have radii related to the mass. The circles are filled with colours according to their escaped ionizing luminosity with the colorbar shown at the bottom. Sink particles with greater mass are plotted on top of those with lower mass to make the former ones more prominent. The time marked at the top-left corner is counted from the end of relaxation.
  Red circles represent stars that are dead and radiation has been shut off.
  The very compact cloud does not have SNe explosion during the duration of the simulation ($\sim 7 \tff \approx 3$~Myr), as all stars live longer than 3~Myr. For the other two less compact clouds, SNe explosions occur when most of the gas is already expelled by radiation. Thus, SNe have little effect on the overall \fesc. For these compact clouds, most of the ionising photons are emitted during middle stage ($3 - 5 \tff$).
  }
  \label{fig:proj_and_Q}
\end{figure*}

\subsection{Calculation of the Ionizing Escape Fraction}
\label{sec:esc}

\begin{figure*}
    \includegraphics[width=\textwidth]{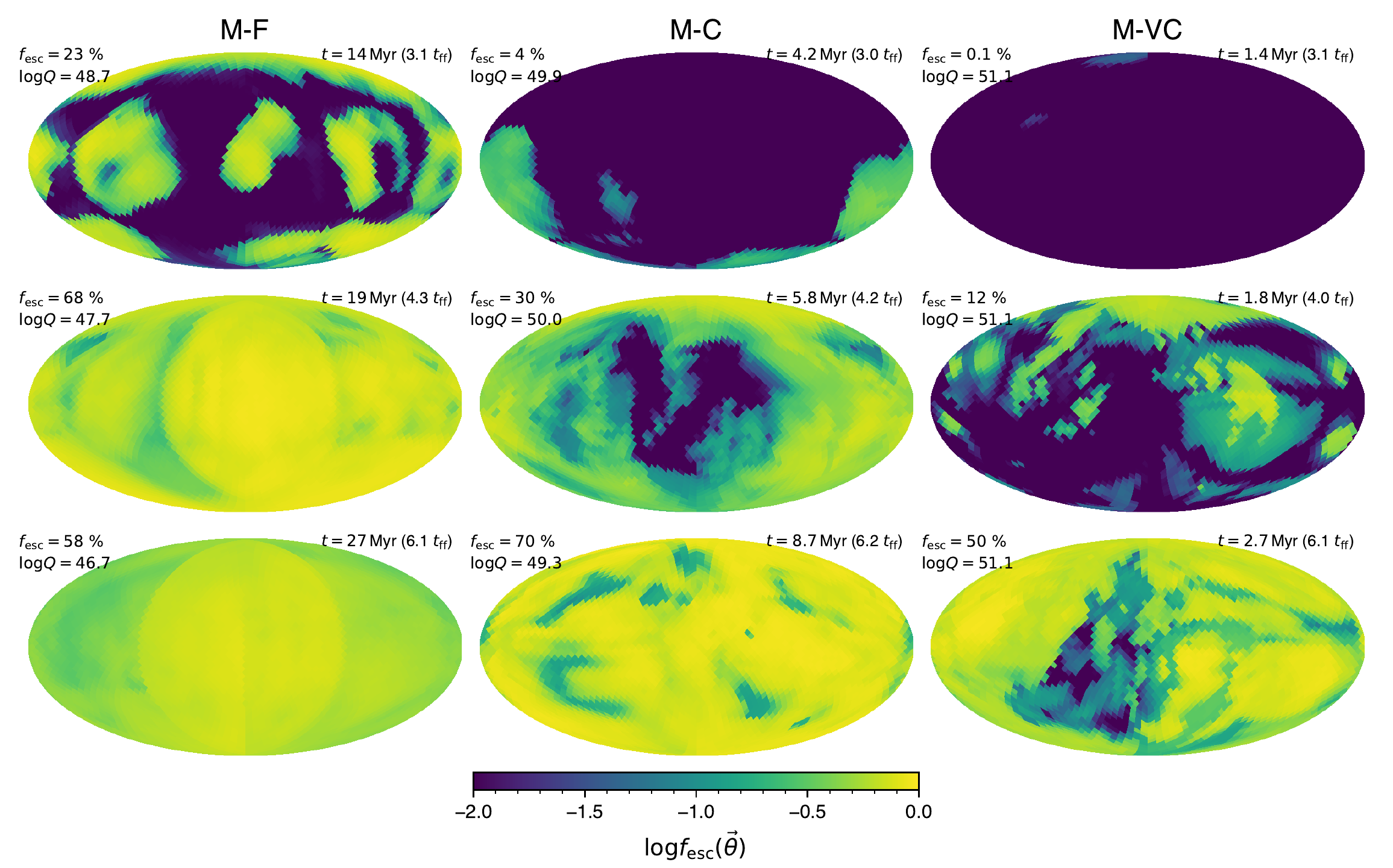}
  \caption{
  Equal-area projection of angular distribution of escape fraction of ionizing photons at three different times (top to bottom) from the Medium mass-Fiducial (M-F), Compact (M-C), and Very Compact (M-VC) clouds, left to right, respectively. 
  The time labelled is the time since relaxation. 
  The escape fraction $f_{esc}(\theta)$, is calculated as a ionizing luminosity weighted average over all stars. The $f_{esc}$ shown in the legend at the top-left corner of each panel, is the average over the whole sky.
  Escaped radiation from star-forming molecular clouds is anisotropic when the cloud is partially ionized. Ionizing chimneys form on part of the sky and expand to the whole sky.
  See the text for how escaped photon emission rate is calculated for individual stars. 
  The hemispherical feature appearing in the bottom-left panel is a numerical artifact that is evident only when \fesc $\sim 1$ and is due to the finite size of the simulation box and the boundary condition.}
  \label{fig:sky}
\end{figure*}

We trace rays from each sink particle and calculate the column density and the optical depth as a function of angular direction. We extract a sphere with radius of the size of the box around each sink particle and pick $12 \times 16^2= 3072 $ directions evenly distributed in the sky using the Mollweide equal-area projection. 
In each direction we implement Monte Carlo integration method to calculate the neutral hydrogen column density by doing random sampling of $\sim 4000$ points in each ray, achieving an accuracy on the escape fraction within $1\%$. The column density is then converted to the escape fraction of ionizing photons in that direction (see Section~\ref{sec:cd_2_ef}). 
The escape fraction from each sink particle is calculated in all directions, then the escape fractions are averaged over all stars, weighting by their ionizing photon luminosity, to get the escape fraction as a function of direction and time, $f_{\rm esc}(\boldsymbol\theta, t)$, from the whole cluster (see Figure~\ref{fig:sky}). We can also define a mean mean escape fraction (averaged over the whole solid angle) from individual sink particles, which is then multiplied by the hydrogen LyC emission rate, $Q$, to get the LyC escaping rate, $Q_{\rm esc}$, as shown in Figure~\ref{fig:proj_and_Q}.

In the calculation of the ionizing escape fraction, the emission from sink particles is shut down after the stellar lifetime, which depends on the mass of the star. We use the equation from \cite{Schaller:1992} as an estimate of the lifetime of a star as a function of its mass, where $M$ is in units of $\msun$:
\begin{equation}
  \label{eq:tstar}
  t_{MS}(M) = \frac{\num{2.5e3} + \num{6.7e2} M^{2.5} +
    M^{4.5}}{\num{3.3e-2} M^{1.5} + \num{3.5e-1} M^{4.5}}~{\rm Myr},
\end{equation}

Note that due to the short lifetime of the clouds after the first star is formed, we do not expect the end of the star's main sequence to significantly affect the dynamical evolution of the simulations (see Section \ref{sec:sims}).

For a subset of simulations we also implement radiative transfer of helium ionizing radiation and helium chemistry (simulations that include this have He escape fractions listed in Table \ref{tab:2}). The calculation of the helium ionizing escape fraction is implemented analogously to hydrogen as explained above. We use fits from \cite{Vacca:1996} for the H-ionizing photon emission rate from individual stars, $Q^{\rm H}$ (or $Q$ for simplicity), and fits from \cite{Schaerer:2002} for $Q^{\rm He}$ and $Q^{\rm He^+}$.

\subsubsection{Conversion from column density to escape fraction}
\label{sec:cd_2_ef}

The neutral hydrogen ionization cross section as a function of frequency is well approximated by a power-law \cite[e.g.][]{Draine:2011}:
\[ \sigma(\nu) \approx \sigma_0 \left( \frac{h\nu}{I_{\rm H}} \right) ^{-3}  \;
\mathrm{for} \; I_{\rm H} < h\nu \lesssim 100I_{\rm H}, \]
where $\sigma_0=6.304\times10^{-18}\,$cm$^{2}$ and $I_{H}=13.6$~eV. 
The escape fraction of photons at a frequency $\nu$ and direction $\boldsymbol \theta$ from a given star is
\begin{equation}\label{eq:fprox}
  f_{esc, \star}(\nu, \boldsymbol\theta) = {\rm e}^{-\tau(\nu, \boldsymbol\theta)} = {\rm e}^{-\sigma(\nu) N_{HI}(\boldsymbol\theta)},
\end{equation}
where $N_{HI}(\boldsymbol\theta)$ is the neutral hydrogen column density from the surface of a star to direction $\boldsymbol\theta$.
If we assume that the stars radiate as perfect black bodies at temperature $T$, then the frequency-averaged escape fraction of hydrogen-ionizing photons is
\begin{equation}\label{eq:ftau}
  f_{\rm esc, \star}(\boldsymbol\theta; T) = \cfrac{\displaystyle\int_{I_{\rm H}}^{\infty} \frac{B_\nu(\nu,T)}{h\nu}
  f_{esc, \star}(\nu, \boldsymbol\theta) \ {\rm d}(h\nu)}{\displaystyle\int_{
  I_{\rm H}}^{\infty} \frac{B_\nu(\nu,T)}{h\nu} \ {\rm d}(h\nu) },
\end{equation}
where $B_\nu(\nu, T) $ is the Planck function. 
The details on how $f_{\rm esc, \star}(\boldsymbol\theta; T)$ behaves for stars with different masses and therefore black-body temperatures is discussed in Appendix~\ref{sec:app1}.

\section{Results}
\label{sec:analyse}

\begin{table*}
  \caption{A summary of results from the analysis of the simulations in Table~\ref{tab:1}. The columns show the number of hydrogen and helium ionizing photons emitted by the star clusters, $S$, and the fraction escaping the clouds, \fesc. 
  \label{tab:2}}
    \centering \def\arraystretch{1.3}
    \begin{tabular}{|c|c|c|c|ccc|cccc|}
      \hline
      Compactness
      & Cloud name 
      & {$m_{\rm gas}/\msun$}
      & {$m_{cl}/\msun$}
      & \multicolumn{3}{c|}{LyC Emission ($\log S/{\rm photons}$)}
      & \multicolumn{4}{c|}{\fesc{} / $\%$} \\
      & & &
      & H 
      & He 
      & He$^+$
      & H LyC
      & H Ly edge
      & He Ly edge
      & He$^+$ Ly edge
      \\
      \hline
      Fiducial & XS-F   & \num{3.2e+03} & \num{3.8e+02} & \gray{62.9} $^{\rm a}$ & \gray{61.6} &
\gray{58.3} &     \gray{53} &     \gray{44} &     \gray{53} &      \gray{0.2} \\
Fiducial & S-F    & \num{1.0e+04} & \num{5.1e+02} & \gray{61.7} & \gray{59.3} &
\gray{57.1} &     \gray{60} &     \gray{53} &     \gray{58} &      \gray{8.2} \\
Fiducial & M-F    & \num{3.2e+04} & \num{1.4e+03} & 63.5 & 62.4 & 59.1 &      8 &    5.2 &    3.7 &    0.063 \\
Fiducial & L-F    & \num{1.0e+05} & \num{5.7e+03} & 64.6 & 63.7 & 61.0 &    2.3 &    1.3 &   0.85 &  1.7e-09 \\
Fiducial & XL-F   & \num{3.2e+05} & \num{2.5e+04} & 65.3 & 64.4 & 61.7 &    1.4 &   0.45 &   0.58 &  9.5e-18 \\
Compact & XS-C   & \num{3.2e+03} & \num{1.0e+02} & \gray{60.5} & \gray{-inf} &
\gray{-inf} &     \gray{92} &     \gray{92} &      \gray{-} &        \gray{-} \\
Compact & S-C    & \num{1.0e+04} & \num{5.3e+02} & 63.3 & 62.2 & 59.0 &     31 &     24 &      - &        - \\
Compact & M-C    & \num{3.2e+04} & \num{3.0e+03} & 64.1 & 63.1 & 59.9 &     23 &     16 &      - &        - \\
Compact & L-C    & \num{1.0e+05} & \num{1.4e+04} & 65.0 & 64.1 & 61.4 &     21 &     14 &      - &        - \\
Compact & L-C-lm & \num{1.0e+05} & \num{3.4e+03} & 64.4 & 63.5 & 60.7 &     44 &     35 &      - &        - \\
Compact & L-C-xlm & \num{1.0e+05} & \num{3.3e+03} & 64.4 & 63.5 & 60.7 &     49 &     43 &      - &        - \\
Very Compact & XXS-VC & \num{1.0e+03} & \num{9.8e+01} & 61.9 & 59.9 & 57.2 &     83 &     79 &     85 &       15 \\
Very Compact & XS-VC  & \num{3.2e+03} & \num{5.1e+02} & 62.8 & 61.4 & 58.2 &     71 &     63 &     71 &      0.2 \\
Very Compact & S-VC   & \num{1.0e+04} & \num{3.2e+03} & 64.4 & 63.5 & 60.8 &     48 &     40 &      - &        - \\
Very Compact & M-VC   & \num{3.2e+04} & \num{1.5e+04} & 65.1 & 64.2 & 61.4 &     35 &     27 &      - &        - \\
Very Compact & L-VC   & \num{1.0e+05} & >\num{2.7e+04} & - & - & - & - & - & - & - \\

      \hline
      \multicolumn{11}{p{16cm}}{$^{\rm a}$ The gray data in this table are from the `XS-F', `S-F', and `XS-C' clouds where the simulation results are less reliable because the SFE is likely overestimated due to missing feedback processes in low-mass stars (see Paper~I).}
  \end{tabular}
\end{table*}

Figure~\ref{fig:proj_and_Q} shows snapshots at times $t \approx 1, 3, 6 ~ \tff$ (top to bottom) for three medium-mass ($\num{3e4}\msun$) cloud simulations with initial mean densities $\overline n_{\rm gas} = \num{2e2}$, $\num{2e3}$, and $\num{2e4} \ {\rm cm}^{-3}$, from left to right, respectively. The free-fall time $t_{ff}$ for these clouds are 4.4, 1.4, and 0.44 Myr, respectively. Each panel shows the density-weighted projection plots of the density (see colorbar on the right of the figure), while the circles show the stars with radii proportional to the cubic root of their masses (see Paper~I for results on the mass function of the stars) and colors representing the number of ionizing photons escaping the cloud per unit time, $Q_{\rm esc}$ (photons/sec), as indicated by the colorbar at the bottom of the figure (see Section~\ref{sec:esc} for details on how $Q_{\rm esc}$ is calculated). Red circles represent stars that are dead and radiation has been shut off. Inspecting the figure, it is clear that the radiation from massive stars that form in the cloud is initially heavily absorbed by the cloud, while at later times, when radiative feedback has blown bubbles and chimneys through which radiation can escape, the radiation from stars can partially escape the cloud.
Massive stars are born deeply embedded in dense clumps, thus their ionising radiation is initially absorbed by the gas and their overall contribution to the total LyC photons is reduced.
A summary of quantitative results for all 16 simulations in Table~\ref{tab:1} is shown in Table~\ref{tab:2}. The meaning of the different quantities in the table is explained in the remainder of this section.
\begin{figure*}
	\centering
	\includegraphics[width=\textwidth]{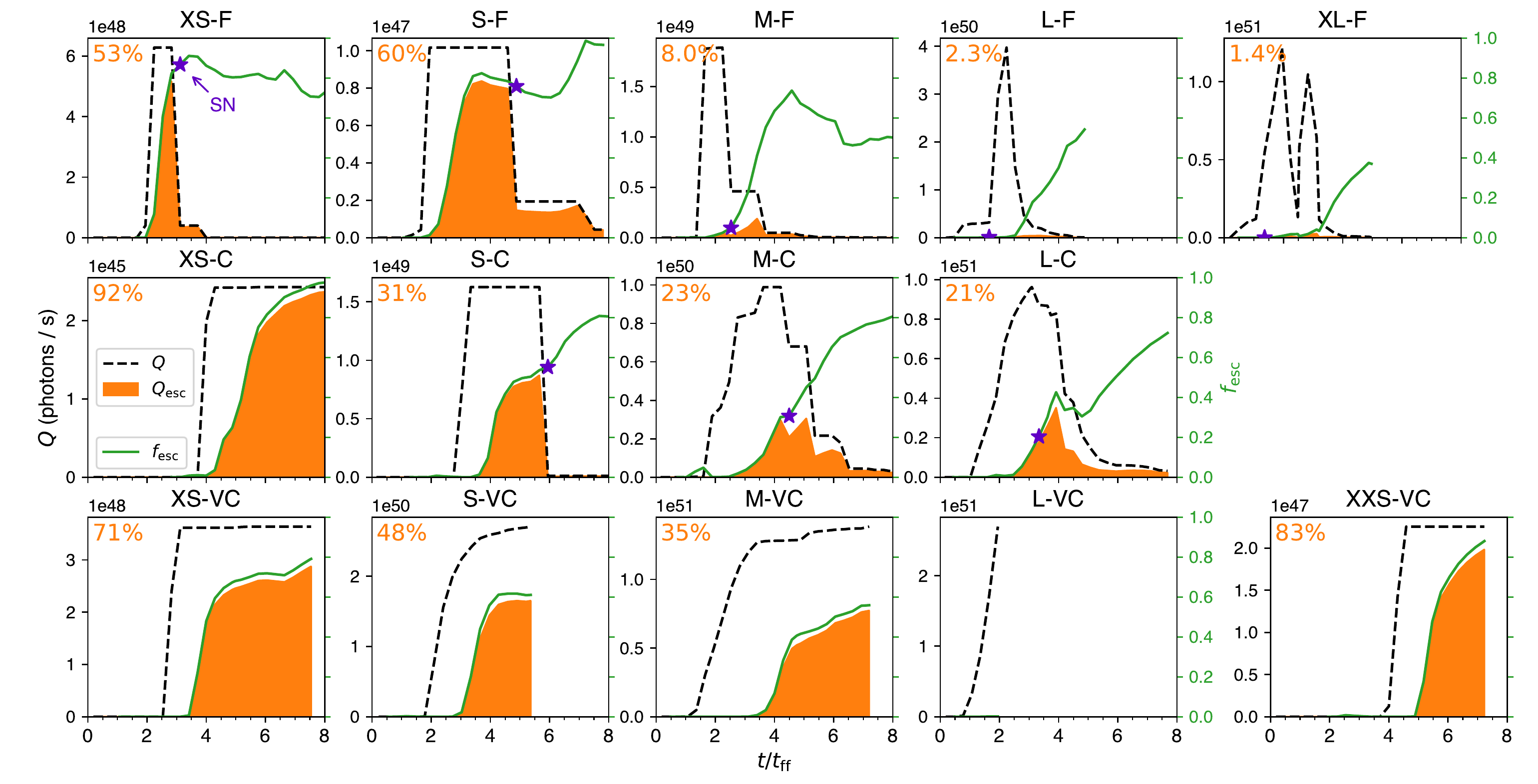}
  \caption{Time evolution of the LyC emission rate ($Q$), escaping rate ($Q_{\rm esc}$), and escaping fraction ($f_{\rm esc} \equiv Q_{\rm esc} / Q$) for our grid of simulations with varying masses (columns) and compactness (rows).
  We notice that in most clouds $f_{\rm esc}(t)$ becomes significant at $3 - 5 t_{\rm ff}$, when most of the volume in the simulation box is ionized. At this time, the Fiducial clouds have a much lower emission rate of ionizing photons ($Q(t)$) with respect to the peak value because the most massive stars in cluster have died, resulting in a low \fesc. The very compact clouds, on the other hand, have a high $Q$ after $3 \tff$, resulting in relatively high \fesc.
  The free-fall times for the clouds in the top (Fiducial), middle (Compact), and bottom panels (Very Compact) are $4.4$, $1.4$, and $0.44$~Myr, respectively.
  The purple stars mark the time when the first SN explosion occurs. Except for the two most massive Fiducial clouds, the first SN explosion happens when $f_{\rm esc}$ is already close to unity and/or when $Q$ has dropped by over an order of magnitude from the maximum, hence in most simulations SN explosions would have little effect on the escape of LyC photons from the cloud.
  }
  \label{fig:Qesc}
\end{figure*}

\subsection{Sky maps of the Escaping Ionizing Radiation}
\label{sec:sky}
Initially, when the radiation starts escaping the cloud (\ie, when the mean value of the escape fraction is small), it does so only in certain directions as illustrated in Figure~\ref{fig:sky} for compact clouds of different masses. The panels are analogous to Figure~\ref{fig:proj_and_Q} (except that the time sequence is chosen differently). Each column shows, for different cloud compactness (density), a time-sequence of sky maps of the leakage of ionizing photons in different directions across the sky using Mollweid projection maps. Columns, from left to right, refer to simulations: M-F, M-C, and M-VC, respectively. Each row refers to a different time: $t=$ 3, 4, and 6 times $t_{\rm ff}$. The clouds start fully neutral and as the first stars form and produce feedback, they start to carve chimneys of ionized gas from where ionizing photons escape. These chimneys then expand and overlap covering larger portions of the sky and finally totally ionizing the whole solid angle. At this time most of the cloud's volume is ionized and $f_{\rm esc}(t)$ is above $10\%$. The neutral fraction in most of the volume is tiny, but due to the large hydrogen column density, the optical depth to LyC photons is typically $\sim 1$, preventing $f_{\rm esc}(t)$ from reaching unity.

However, for the small and medium mass clouds, by the time most of the radiation escapes isotropically, the emission rate of ionizing photons is small because all massive stars have died. In addition, if we consider that these molecular clouds are embedded into galactic disks, the high $f_{esc}(\boldsymbol\theta)$ channels will be randomly oriented with respect to the disk plane, further reducing \fesc and increasing the anisotropic leaking of ionizing radiation.

The escape fraction is anisotropic at early times when most of the radiation from massive stars is emitted. Later, when the leakage of ionising radiation become more isotropic, massive stars, which dominate the ionizing radiation emission, start to die. In the next section we will average the rates of ionizing radiation emitted, $Q$, and escaping $Q_{\rm esc}$, over the whole solid angle and analyse in detail the time evolution of these quantities and calculate the instantaneous escape fraction defined as $f_{\rm esc}(t) \equiv Q_{\rm esc}(t) / Q(t)$. We will see that unless $\fesc \simgt 50\%$ (averaged over the whole sky and over time), the radiation escaping a star cluster is highly anisotropic.

\subsection{Time Evolution of the Sky-Averaged Escape Fraction}
\label{sec:time}

Figure~\ref{fig:Qesc} shows the emission rate of hydrogen-ionizing photon, $Q$ (dashed lines), the portion that escapes from the cloud, $Q_{\rm esc}$ (shaded regions), and the instantaneous escape fraction, $f_{\rm esc}(t) \equiv Q_{\rm esc} / Q$ (solid lines) as a function of time for all our simulations with solar metallicity. 

The stellar lifetime is calculated as the main-sequence lifetime \cite{Schaller:1992} of a star with mass $40\%$ of the sink mass (see Paper~I). Radiation from a star is turned off after the star is dead. As a result, there is a sharp drop of $Q(t)$, thus $Q_{\rm esc}(t)$, after about 3-5~Myr from the beginning of star formation due to the death of the most massive stars in the cluster. Some of our simulations have not been run sufficiently long for all massive stars to die, as we stop the simulations after roughly $6t_{\rm ff}$, when feedback has shut down star formation in the cloud. In all our simulations, except for the `L-VC' run, the SFE reaches its maximum long before the end of the simulation, therefore we are able to extrapolate $Q(t)$ beyond the end of the simulation. We calculate the total number of ionizing photons emitted by the star cluster $S = \int^{t_{end}}_0 Q(t^\prime) \mathop{dt^\prime}$, and the total number of ionising photons that escape the molecular cloud, $S_{\rm esc} = \int^{t_{end}}_0 Q_{\rm esc}(t^\prime) \mathop{dt^\prime}$, where $t_{end}$ is chosen to be the end of the simulation or a sufficiently long time after the end of the simulation such that all massive stars in the simulation have died. We define a time-averaged total escape fraction of ionizing photons as \fesc$\equiv S_{\rm esc} / S$, which is shown in the top-left corner of each panel in Figure~\ref{fig:Qesc}.
\begin{figure*}
    \centering
    \includegraphics[width=\columnwidth]{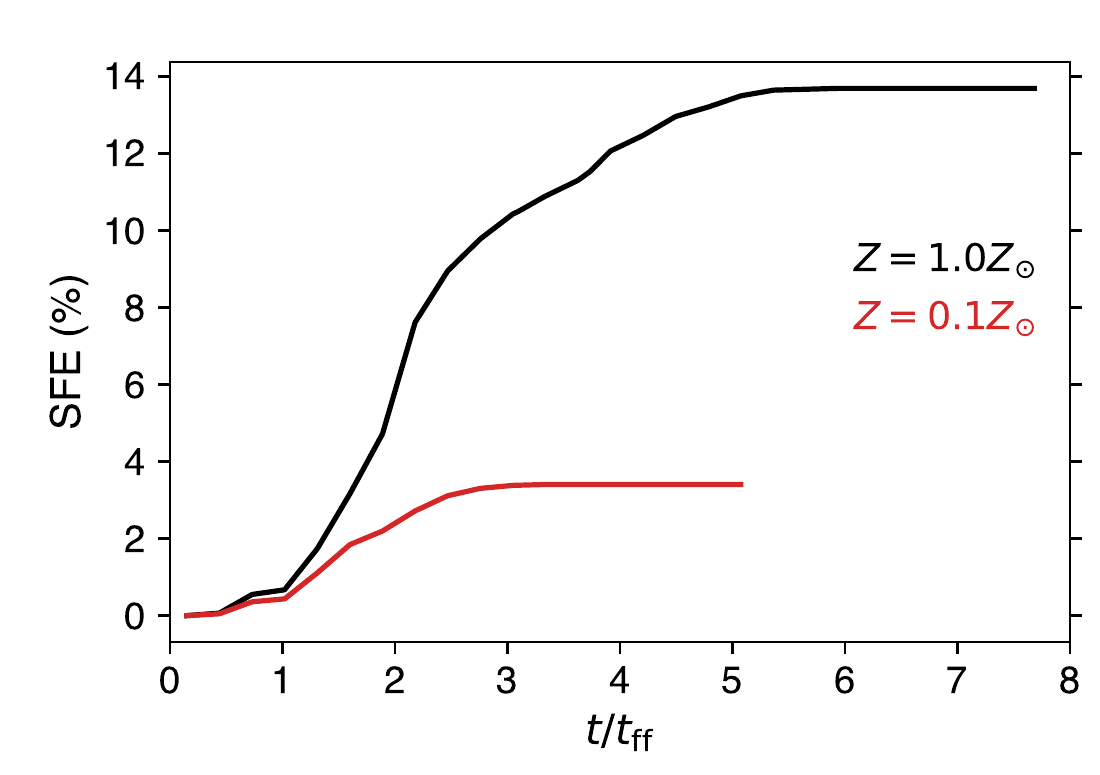}
    \includegraphics[width=\columnwidth]{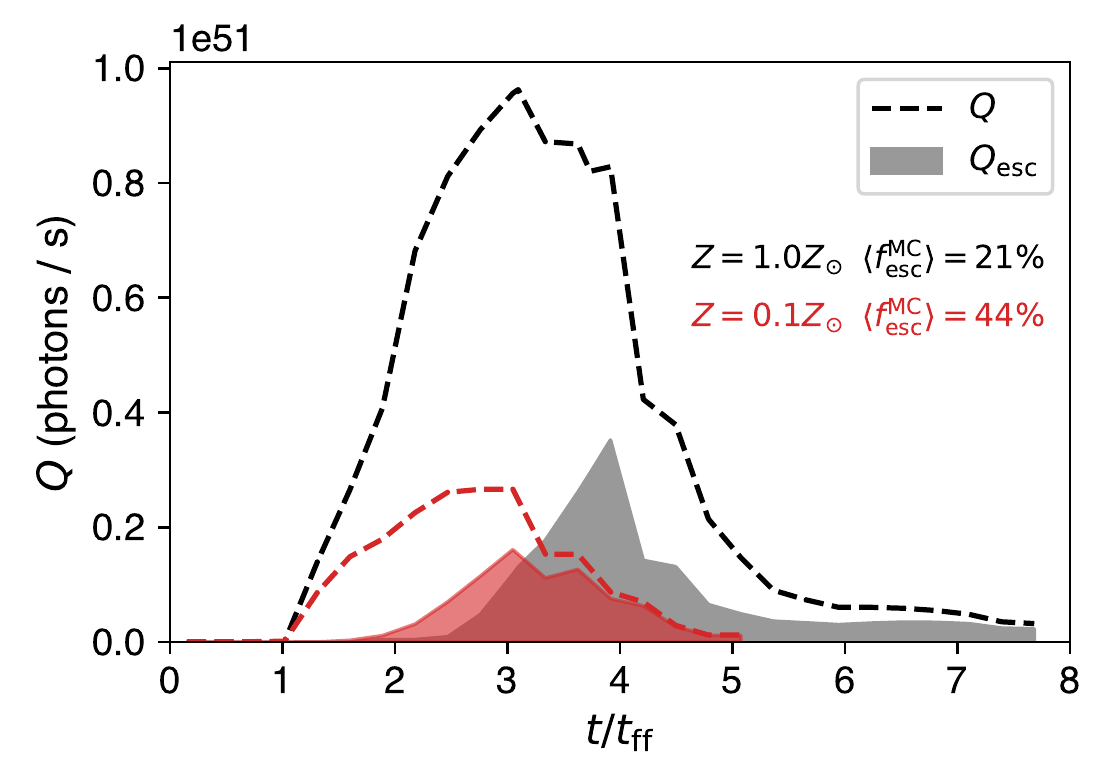}
    \caption{
    {\it (Left).} Time evolution of the SFE for the large Compact (L-C) run with solar metallicity (black line) and $Z=1/10$~Z$_\odot$ (red line). {\it (Right).} Hydrogen ionizing-photon emission rate (dashed lines) and escaping rate (shaded area) for the same simulations as in the left panel. The lower-metallicity run (red lines) has $\sim 3$ times lower photon emission rate $Q$ due to the lower SFE. The stronger stellar feedback in the lower-metallicity cloud clears out the gas in less than $3 \tff$, when star formation is quenched and the escape fraction approaches unity as indicated by the convergence of the $Q_{\rm esc}$ and $Q$ curves, resulting in higher \fesc but substantially lower number of total escaped ionizing photons.
    }
    \label{fig:Q-met2}
\end{figure*}
The figure shows that $f_{\rm esc}(t)$ is practically zero at the time star formation begins when massive stars start emitting ionising radiation. After a time delay $f_{\rm esc}(t)$ increases almost linearly with time and in several simulations it reaches a roughly constant value as a function of time  after $t \sim 5\tff$. This is the time when the bulk of the gas is blown away by radiation feedback and the remaining gas is mostly ionized (see Paper~I). For the simulations in which we do not have a sufficiently long time evolution to measure $f_{\rm esc}$(t) until the time all massive stars have left the main sequence, we assume that $f_{\rm esc}(t)$ maintains the same value found at the end of the simulation and we calculate $Q_{\rm esc}(t)$ from $Q(t)$ and $f_{\rm esc}(t)$ up to the time when all massive stars are dead. 
We will further discuss the results for the integrated ionising photon emission in Section~\ref{sec:ef}.

Mechanical energy and metal enrichment from SN explosions is not included in our simulations. We compensate for the missing feedback by not shutting down UV radiation after the star dies (see Paper~I). Note, however, that in the calculation of \fesc we consider realistic lifetimes of massive stars. 
As shown in Figure~\ref{fig:Qesc} (star symbols), SNe explosions happen typically either when $Q$ is already small and $\fesc \sim 1$, or after the end of the simulation. The only exception is the two most massive fiducial clouds. 
For the Compact and Very Compact clouds as well as the less massive fiducial clouds, both the star formation time scale and feedback time scale (related to the sound crossing time) are shorter than the first SN explosion time ($\sim 3$~Myr).
Therefore, we may have underestimated \fesc in the two most massive fiducial clouds, although enrichment from SN may also reduce \fesc if dust is produced on sufficiently short time scale.

\subsubsection{Effects of Gas Metallicity}
\label{subsec:met}

Figure~\ref{fig:Q-met2} compares two simulations of the L-C cloud, with the only difference being the gas metallicity which affects the cooling of the gas.
For a given cloud mass and density, lowering the gas metallicity increases \fesc, even though here we do not consider dust opacity. In Paper~I we found that for gas metallicity $Z<0.1$~Z$_\odot$, the SFE is reduced by a factor of $\sim 5$ due to more efficient UV feedback caused by the higher temperature and pressure inside \HII regions, but we do not observe a dependence of the IMF on the metallicity. From Figure~\ref{fig:Q-met2} we can see that the peak value of $Q(t)$ for the lower metallicity simulation is reduced with respect to the solar metallicity case by a factor of 4 due to the lower SFE. However, the timescale over which $f_{esc}$ increases from 0 to some value of order unity is shorter with decreasing metallicity, suggesting a faster destruction of the cloud due to a more efficient feedback, in agreement with what we found in Paper~I. We will investigate quantitatively the dependence of \fesc on feedback time scale in Section~\ref{sec:model} with an analytic model.

\subsection{Time-Averaged Escape Fraction \texorpdfstring{\fesc{}}{fesc}}
\label{sec:ef}

\begin{figure}
    \centering
    \includegraphics[width=\columnwidth]{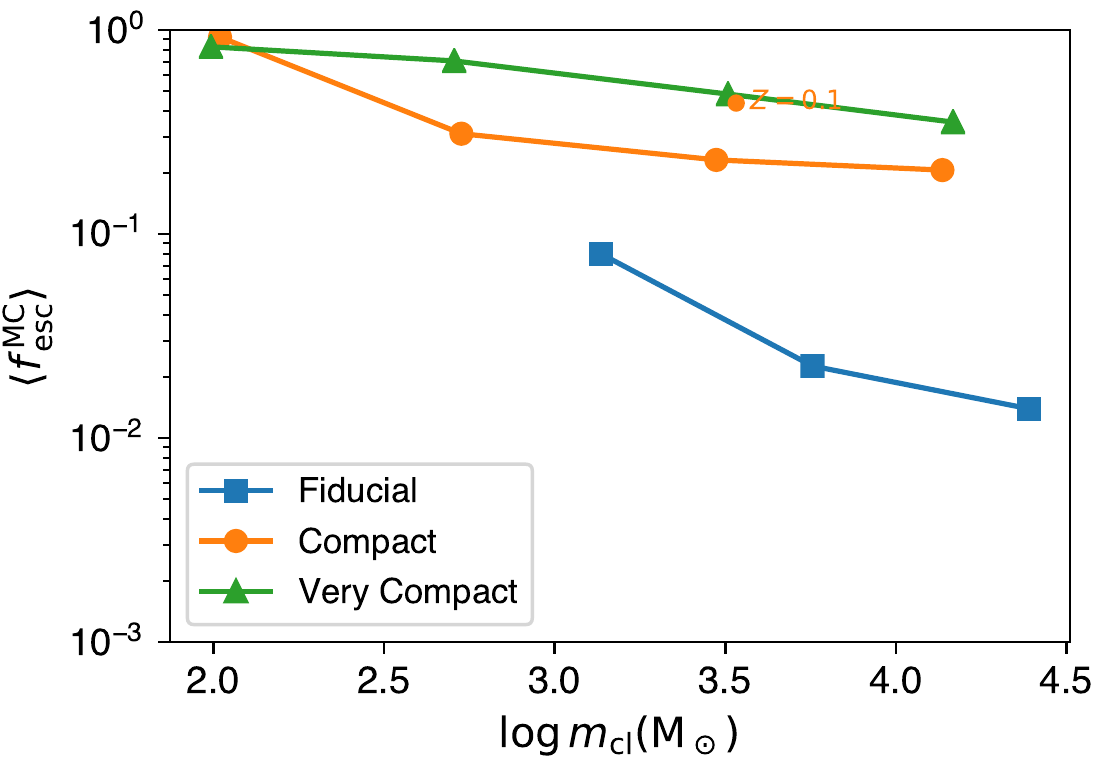}
    \caption{
    The total escape fraction of ionizing photons \fesc = $S_{\rm esc}/S$. The blue, orange, and green lines in both panels connect clouds with same density to guide our eyes. The low-mass clouds have high \fesc due to lower mass stars dominating UV radiation (and lower mass stars live longer)
    }
    \label{fig:fesc}
\end{figure}
Figure~\ref{fig:fesc} summarizes the final result for the escape fraction for all our simulations, showing \fesc $\equiv S_{esc}/S$ as a function of the mass of the star cluster, $m_{\rm cl}$, for different molecular cloud compactness (as shown in the legend). 
The two least massive fiducial clouds are removed from the analysis because we believe that the SFE of these simulations is overestimated due to missing physics (\ie, IR feedback, that is not included in these simulations, becomes significant in this regime. See Paper~I for more explanation).
We find that \fesc increases with decreasing mass of the cluster and with increasing compactness. We also find a strong dependence of \fesc on the gas metallicity.

As we decrease the gas metallicity, the typical pressure inside \HII regions increases. Therefore the feedback becomes stronger, leading to an increases of \fesc, but also a reduction of the SFE. Therefore, the total number of escaped LyC photons decreases with decreasing metallicity, because of  the reduced SFE.

\begin{figure}
    \centering
    \includegraphics[width=\columnwidth]{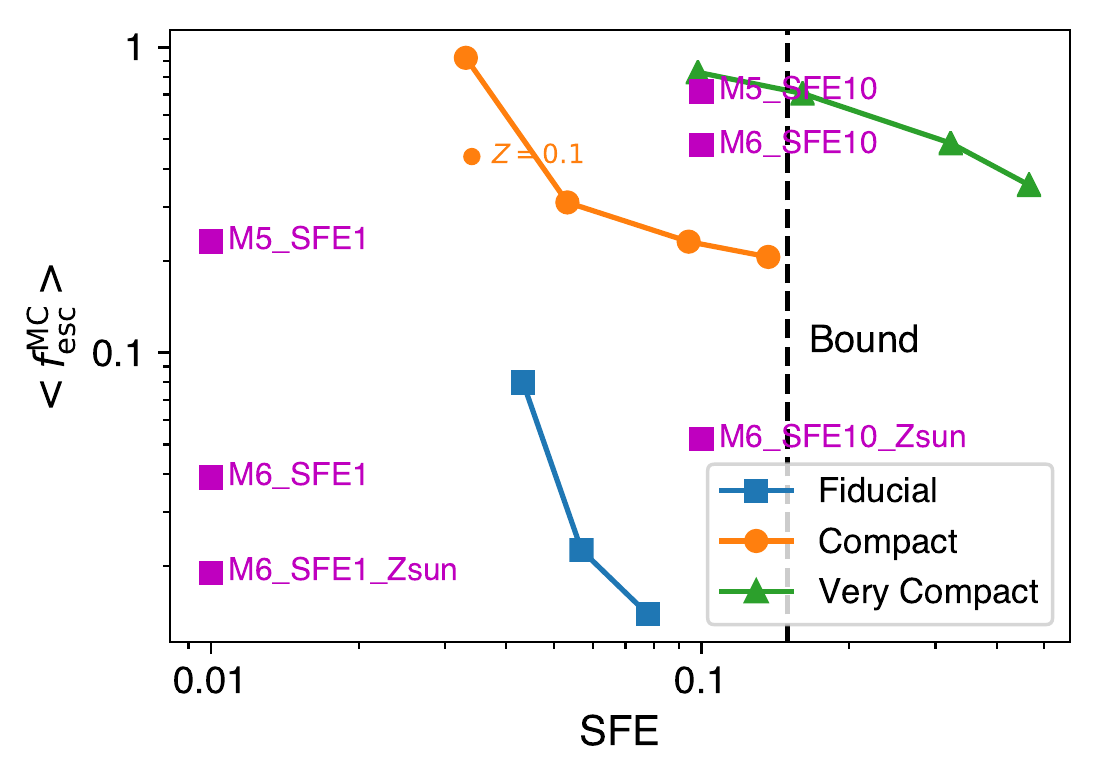}
    \caption{\fesc plotted against SFE for 11 out of our 16 simulations. Magenta squares are data from \protect\cite{Kimm:2019}. In the labels the number after `M' refers to $\log_{10}(m_{\rm \scriptscriptstyle MC}/\msun)$, and the number after `SFE' is the SFE in per cent. The metallicity is $0.1 \, Z_\odot$ unless otherwise specified.}
    \label{fig:fesc_sfe}
\end{figure}
Figure~\ref{fig:fesc_sfe} shows \fesc as a function of the SFE for 12 out of our 16 simulations. For comparison, results from \cite{Kimm:2019} are plotted as purple squares.
The methodology in the simulations by \cite{Kimm:2019} is rather different from ours, because star formation is not modelled self-consistently but rather a fixed SFE (of $1\%$ or $10\%$) is assumed and stars placed at the center of the cloud inject energy and radiation according to a pre-computed stellar population. They assume gas clouds of fixed density, similar to our fiducial case, and explore masses of $10^5$~M$_\odot$ and $10^6$~M$_\odot$ and metallicities of $0.1$ solar and solar metallicity.

In Paper~I we have shown that there is a tight positive correlation between the SFE and $m_{\rm cl}$. Therefore in our simulations \fesc decreases with increasing cloud mass and therefore with increasing SFE. 
The results for gas at solar metallicity and the dependence of \fesc on the cloud mass are in qualitative agreement with \cite{Kimm:2019}, as well as the significant increase of \fesc as the gas metallcity is reduced with respect to the solar value.

For the fiducial clouds, with densities typical of star forming regions in the local Universe, \fesc is extremely small: going from $\fesc \sim 8\%$ for star clusters of $10^3$~M$_\odot$, to  $1.4\%$ for clusters of $3\times 10^4$~M$_\odot$. Clearly if high-redshift star clusters had the same properties as today's ones, their \fesc would be too low to contribute significantly to the reionization process.
However, for our compact and very compact clouds, we find higher values of \fesc: ranging from \fesc$>50\%$ for clusters of mass $<500$~M$_\odot$, to $20\%$ (compact) and $35\%$ (very compact) for star clusters with masses $\sim 2\times 10^4$~M$_\odot$.

We emphasize that \fesc we are reporting in this work is an upper limit for \fescg from galaxies. Here we are simulating the escape fraction just from the molecular clouds, without including a likely further reduction of \fescg due to absorption of ionising radiation by the ISM in the galaxy. We also do not include the effect of dust. Therefore, even for compact clouds, \fesc is already quite close to the average value required for reionization, which is an interesting result in order to understand the nature of the sources of reionization.
\begin{figure}
  \centering
  \includegraphics[width=\columnwidth]{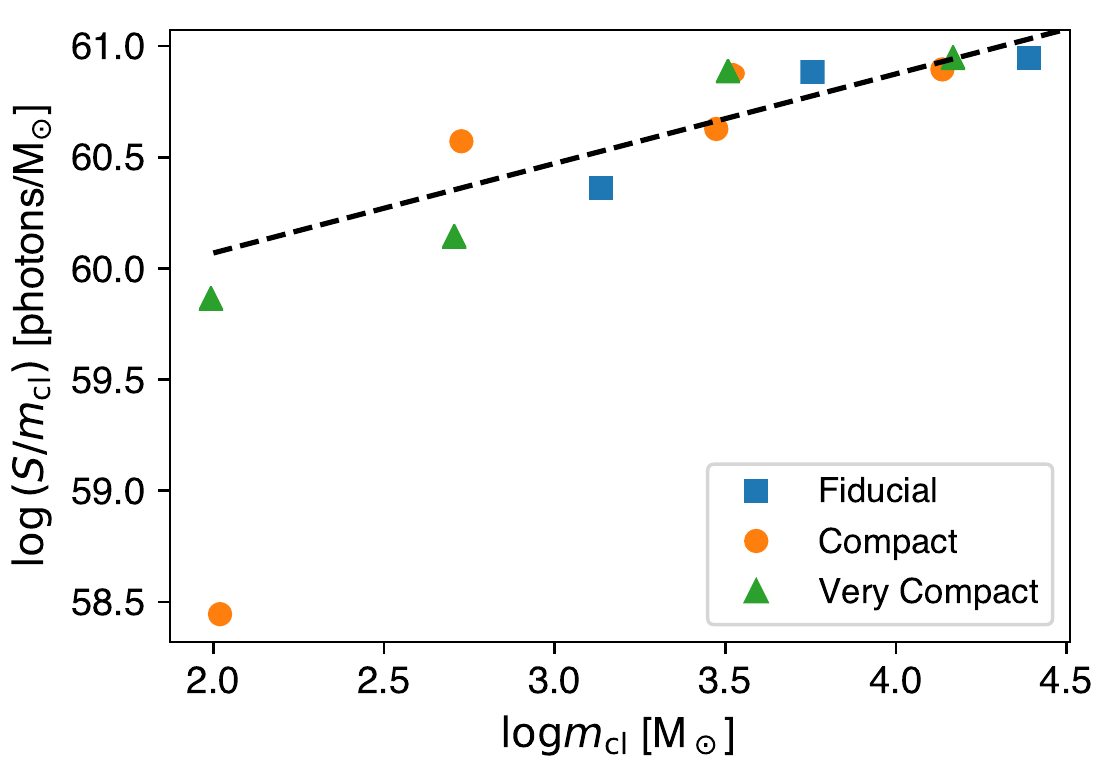}
  \includegraphics[width=\columnwidth]{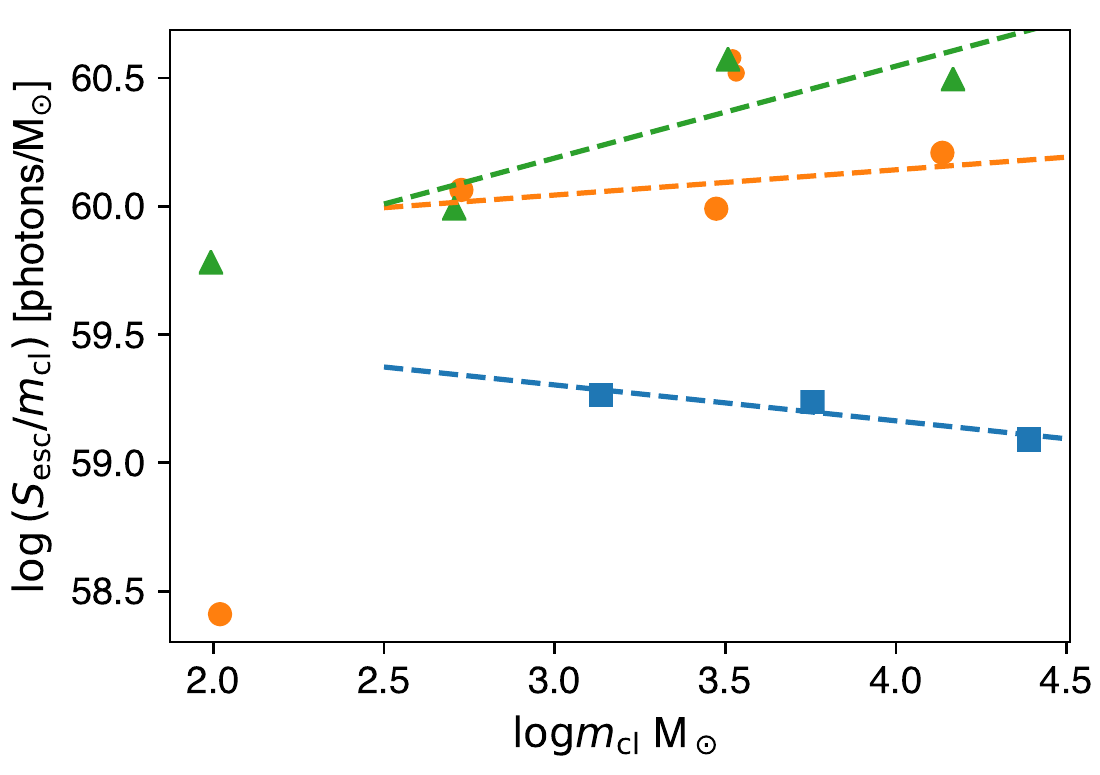}
  \caption{
  Time-integrated number of hydrogen-ionizing photons per unit star cluster mass emitted (top) and escaping the cloud (bottom) as a function of the star cluster mass and for gas cloud densities as in the legend. The relation between $S/m_{cl}$ and $m_{cl}$ is a tight power-law function with a slope $\sim 0.4$, independent of the density of the initial cloud, as expected. For clusters forming in molecular clouds with same initial density, the relationship between $S_{esc}/m_{cl}$ and $m_{cl}$ is also well approximated by a power-law with negative slope for the fiducial clouds (local Universe clouds) and increasing positive slope with increasing cloud compactness.
  }
  \label{fig:Sesc}
\end{figure}

A complementary way to characterise the ionising radiation escaping molecular clouds is in terms of $S_{esc}$ or $Q_{esc}$. Since more massive star clusters emit more ionizing radiation per unit mass, these quantities show more directly the relative importance of clusters with different mass  to the total ionising photons escaping a galaxy. The top panel in Figure~\ref{fig:Sesc} shows the total number of ionizing photons emitted by the cluster per unit mass, $S/m_{cl}$, over its lifetime as a function of the mass of the star cluster for all the simulations in Table~\ref{tab:1}. The dashed line shows a power-law fit to $S/m_{cl}$ as a function of $m_{cl}$, excluding the two data points with $m_{cl} \simlt 300$~M$_\odot$: 
\begin{equation}\label{eq:smcl}
    \frac{S}{m_{cl}} = \num{1.2e60} \left( \frac{m_{cl}}{100\msun} \right)^{0.4}.
\end{equation}
We exclude from the fit star clusters with mass below $300\msun$ because for small mass clusters the scatter of $S$ becomes very large due to sparse sampling of massive stars in small clouds (see Figure~7 in Paper~I). We can roughly understand the $0.4$ slope of the power-law fit by assuming that the most massive star in a cluster dominates the emission of ionising radiation. In Paper~I we found that the most massive star in the cluster has a mass $M_{max} \propto m_{cl}^{0.66}$, and for stellar masses $M \simgt 30\msun$, $Q(M) \propto M^{1.9}$ with a lifetime on the main sequence $t_{MS}(M)$ nearly constant as a function of mass. Thus, we get $S \propto Q(M_{\rm max}) \propto m_{cl}^{1.3}$ and $S/m_{\rm cl} \propto m_{cl}^{0.3}$, which is close to the exponent in Eq.~(\ref{eq:smcl}).
We will show later that the most massive star in the cluster typically contributes a fraction $25\%$ to $95\%$ of all the emitted ionising photons.

The bottom panel in Figure~\ref{fig:Sesc} shows the total number of ionising photons escaping the cloud per unit mass, $S_{esc}/m_{cl}$, as a function of the cluster mass for the same simulations as in the top panel. The dashed lines show power-law fits
\begin{equation}
\frac{S_{esc}}{m_{cl}} = E \left(\frac{m_{cl}}{100~M_\odot}\right)^{\alpha},\\
\label{eq:s_esc}
\end{equation}
where $E=\num{2.8e+59}$, $\num{8.8e+59}$, and $\num{6.8e59} \msun^{-1}$ and $\alpha=-0.1, 0.1, 0.4$ for the fiducial, compact and very compact clouds, respectively.
The figure shows that for clouds in the local Universe (fiducial clouds) and for compact clouds, the number of escaping ionising photons per unit mass ($S_{esc}/m_{cl}$) is nearly constant with increasing cluster mass, while for very compact clouds $S_{esc}/m_{cl}$ increases with increasing cluster mass. We will see in Section~\ref{sec:ob} that this trend is reflected in the total number of escaping ionising photons integrated over the observed (in the local Universe) star cluster mass function.

Combining Eqs.~(\ref{eq:smcl}) and (\ref{eq:s_esc}), the power-law fitting function for the escape fraction is
\begin{equation}
\langle f_{esc}\rangle =F \left(\frac{m_{cl}}{100~M_\odot}\right)^\beta,
\label{eq:fescfit}
\end{equation}
where the power-law slopes are $\beta=\alpha-0.4=-0.5, -0.3, 0.0$ and normalizations $F=0.23, 0.73, 0.57$ for the fiducial, compact and very compact clouds, respectively.

\begin{figure}
    \centering
    \includegraphics[width=\columnwidth]{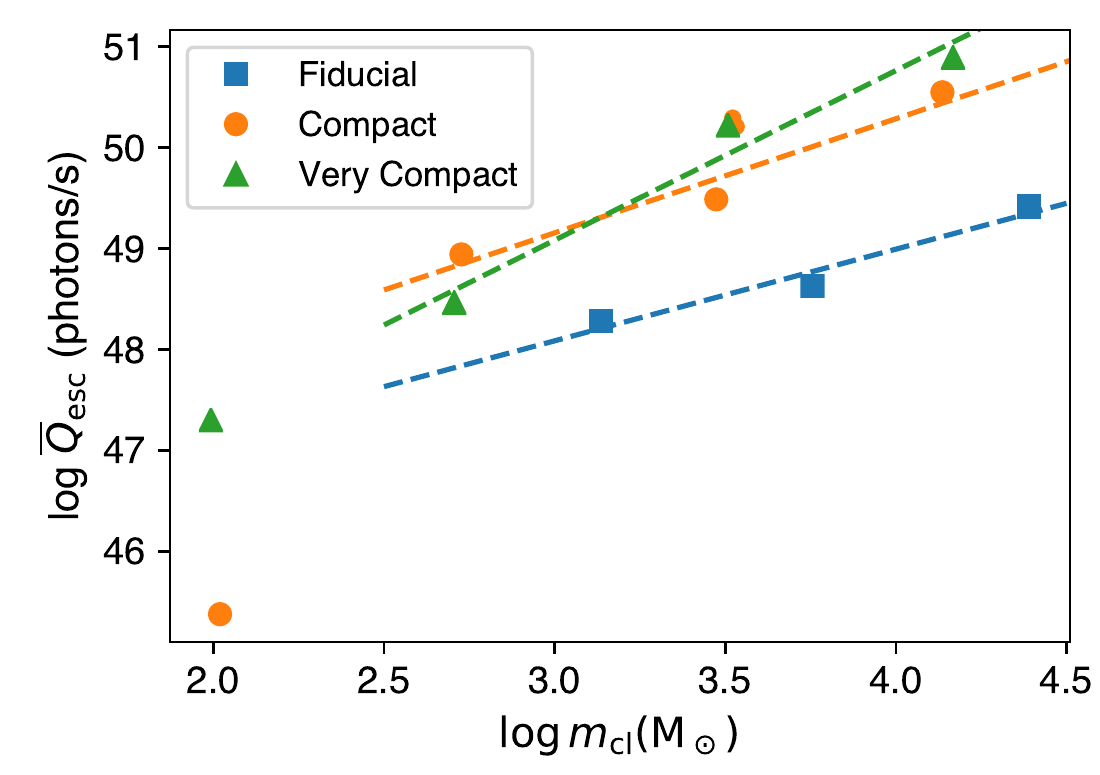}
    \includegraphics[width=\columnwidth]{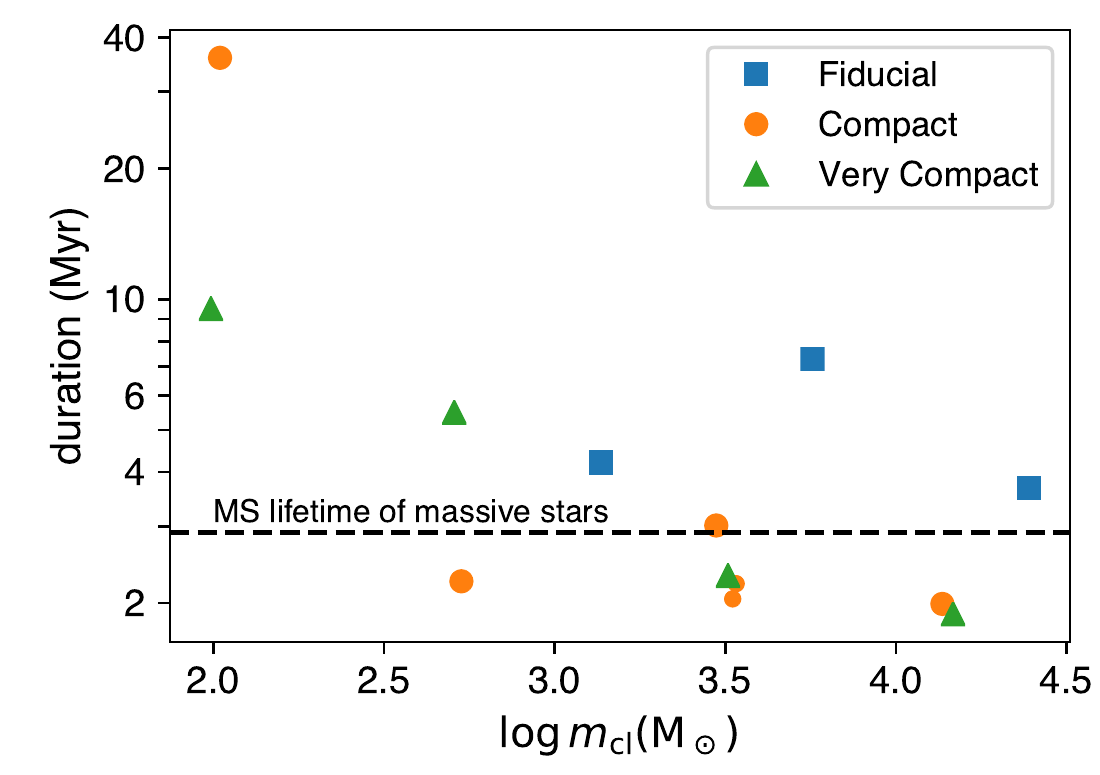}
    \caption{Top: Mean escaped ionizing-photon emission rate $Q_{esc}$ as a function of cluster mass $m_{cl}$. Power-law fits to each group of data is shown as dashed lines with corresponding colors. The slopes are 1.7, 2.2, and 0.4 for the VC, C, and F clouds, respectively.
    Bottom: Duration of ionizing-photon escaping.
    }
    \label{fig:ef_hw}
\end{figure}

\begin{figure*}
    \centering
    \includegraphics[width=\textwidth]{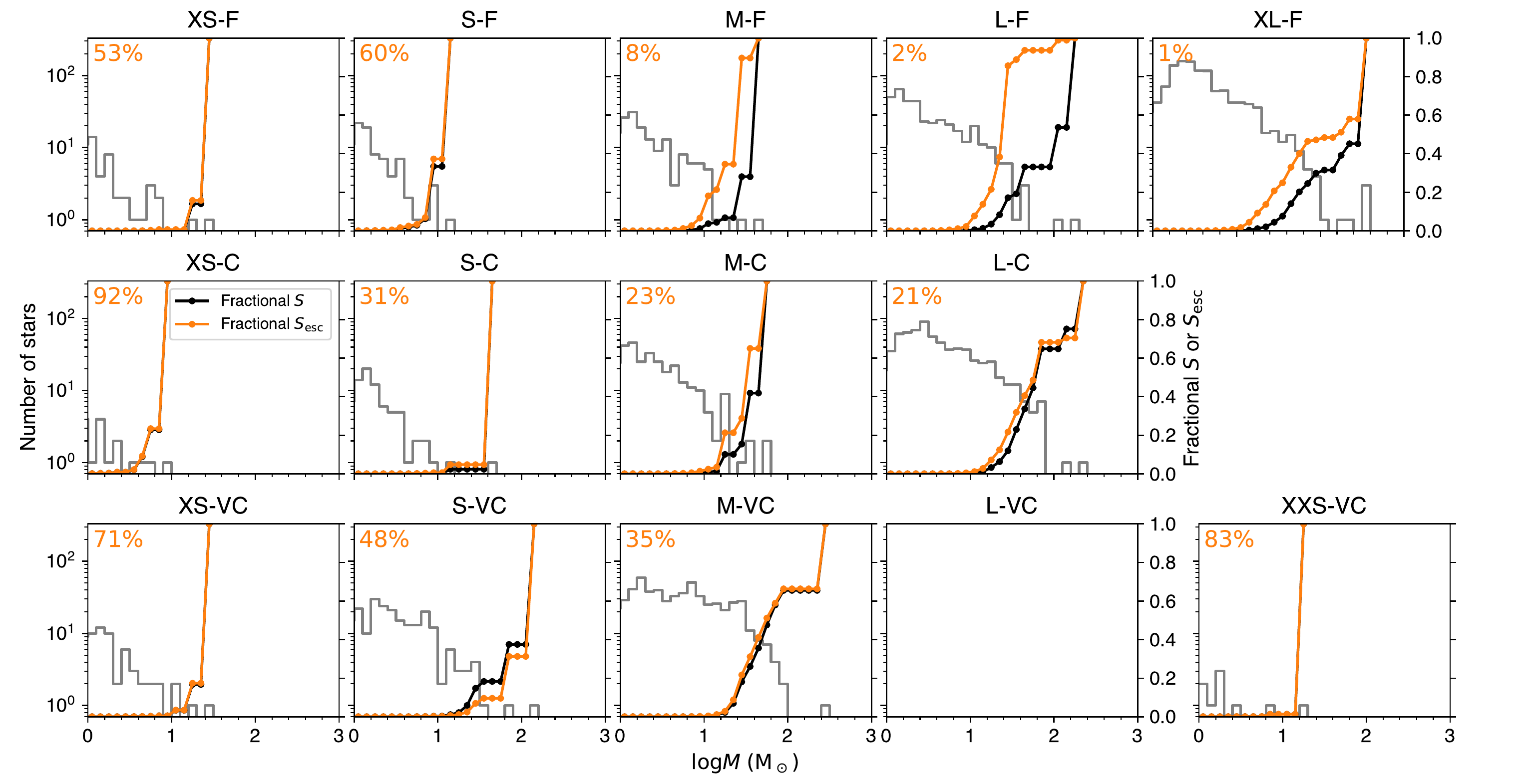}
    \caption{Fractional cumulative radiation emission (black) and escaping (orange) as a function of the mass of the star. The gray histogram shows the numbers of stars per log bin. 
    The black lines show that, although there are on average only a few massive stars in clusters, they dominate the emission of ionizing radiation. Inspecting the orange lines, we see that, except for the two most massive fiducial runs (L-F and XL-F), the same is true for the total escaped radiation from the cluster. In the massive fiducial clouds, very massive stars live shorter than a free-fall time ($\sim 4$~Myr) and die before the gas is ionized and radiation can escape. This also results in lower ($\lesssim 5\%$) \fesc (the numbers on the top-left corner of each panel.)}
    \label{fig:inhom}
\end{figure*}

In cosmological simulations and analytic models, the sources of ionising radiation are typically modelled as sub-grid physics in terms of the mean ionising photon escape rate $\overline Q_{esc}$ during the UV burst, and the duration of the ionising burst $t_{esc}$. The duration of the burst and the anisotropy of the radiation escaping galaxies actually plays an important role in determining the photon-budged for completing IGM reionization and the topology of reionization \citep{Hartley:2016}. These quantities for star forming molecular clouds are shown in Figure~\ref{fig:ef_hw} as a function of the stellar cluster mass $m_{cl}$, where we approximate $\overline Q_{esc}$ as the peak value of $Q_{esc}(t)$ and define $t_{esc} \equiv S_{esc}/\overline Q_{esc}$.

The dashed lines are power-law fits to the data. We find 
\begin{equation}
\overline Q_{esc} = Q_0 \left(\frac{m_{cl}}{100~M_\odot}\right)^{\gamma}\\
\label{eq:Q_esc}
\end{equation}
where $\gamma=0.9, 1.1, 1.7$ and $Q_0=\num{1.5e+47}$, $\num{1.1e+48}$, $\num{2.5e+47}$~s$^{-1}$, for the fiducial, compact and very compact clouds, respectively. 
For the local Universe clouds (fiducial case), $\overline Q_{esc} \sim 10^{48} - 3 \times 10^{49}$~s$^{-1}$ in the range $m_{cl} \sim 10^3-3\times 10^4$~M$_\odot$, increasing nearly linearly with increasing cluster mass. We have also noticed that, if we consider $\overline Q_{esc}$ of radiation at the hydrogen ionization edge ($13.6$~eV) rather than the weighted mean over the stellar spectrum (see Appendix~\ref{sec:app1}), we find that $\overline Q_{esc}$ is nearly constant as a function of the cluster mass, in good agreement with \cite{Dale:2014}. For very compact star clusters, however, the dependence on the mass is quite strong: $\overline  Q_{esc} \sim 5\times 10^{48}$~s$^{-1}$ for $m_{cl} \sim 500$~M$_\odot$, but increases to $10^{51}$~s$^{-1}$ for $m_{cl} \sim 20,000$~M$_\odot$. For the very compact and, to some extent, for the compact clouds, the duration of the burst of ionising radiation escaping the molecular cloud reflects the duration of the emitted radiation, that is roughly the lifetime of the most massive star formed in the cluster (\ie, $t_{burst} \sim t_{uv} \approx t_{MS}(M_{max})$), although the emitted radiation is partially absorbed by the gas cloud. Hence, for small mass clusters the duration of the burst is longer: increasing from 2~Myr for $m_{cl} \sim 10^4 $~M$_\odot$ to $10$~Myr for $m_{cl}\sim 100$~M$_\odot$.  However, this trend with the cluster mass is not observed for the two most massive fiducial clouds, for which $t_{burst}\sim 7$~Myr, about twice as large as the duration of the emitted burst of ionising radiation $t_{uv} \sim t_{MS}(M_{max})$. The reason for why the effective timescales of the emitted and escaping radiation differ from each other, can be found inspecting Figure~\ref{fig:inhom} for those two clusters. 
For massive clusters, especially when \fesc is very small, not only the most massive star, but also stars with $M \sim 10-20$~M$_\odot$ contribute to $Q_{esc}$. Hence, the effective timescale for the escaping radiation can be longer than the effective timescale when most of the ionising radiation is emitted.

\subsection{Escape Fraction of Helium Ionising Photons}

Having discussed the emission rate of hydrogen-ionising photons, we explore another group of photons that ionize He and He$^+$. We enable the emission of these photons from sink particles in a subset of our simulations (the fiducial simulations plus the least massive compact and very compact runs). Massive stars with non-zero metallicity do not emit \HeII ionising photons with energy $>54$~eV, hence we will not consider this energy bin\footnote{Wolf-Rayet stars actually emit some \HeII ionising radiation, but so far we have not included these type of stars in our simulations.}.

We find that in all the simulations in which we include photon bins that ionize He, the escape fraction of HeI-ionising photons is nearly identical to that of HI-ionizing photons, with the only exceptions of the three most massive fiducial clouds where the \fesc for HeI is lower by a factor of 2 -- 3.

We interpret this result arguing that the sizes of \HII and He$^+$ ionization fronts are comparable around the sources that dominate the emission of ionising radiation. The radius of the ionization front can be estimated using the Str\"omgren radius equation:
\begin{equation}
  R_{S0}^{i} \equiv \left( \frac{3 Q^i}{4 \pi {n_i}^2 \alpha_B^i} \right)^{1/3},
\end{equation}
with $i$ being H or He$^+$.
At $10^4$~K, the case-B recombination rate, $\alpha_B^{\rm He^+}$, is about $1.9$ times higher than that of hydrogen.
With a He abundance ratio $n_{He}/n_H = (\mu-1)/(4-\mu) \approx 0.154$, where $\mu = 1.4$ is the mean atomic weight of the gas in our simulations, the \HeII front is larger or equals the radius of the \HII I-front when the hardness of the spectrum, $Q^{\rm He}/Q^{\rm H}$, is greater than $0.29$. Hot O stars have spectrum hardness close to or above this critical value, therefore around massive stars, which dominate the ionizing radiation, the \ion{He}{i}-front is slightly larger than the H ionization front. Therefore, we expect that \fesc for He-ionizing photons is close to or slightly larger than that of H-ionizing photons. This expectation is supported by the analysis of all our simulations that include radiation transfer in the He-ionizing frequency bins (see Table~\ref{tab:2} as well as left panel of Figure~\ref{fig:fesc_ev}).

\subsection{Absorption by Dust}

It is well known that dust may contribute significantly to the absorption of ionizing radiation \citep[\eg,][]{Weingartner:2001}. 
In this section we estimate the effect of dust absorption on the escape fraction of LyC photons by adopting the dust extinction parameterization for the Small Magellanic Clouds (SMC) in \cite{Gnedin2008}, which is based on \cite{Pei1992} and \cite{Weingartner:2001}.
When dust absorption is included, the escape fraction in each direction is 
\begin{equation}
    f_{\rm esc}(\nu, \boldsymbol\theta) = f_{\rm esc, gas} {\rm e}^{-\tau_d(\nu, \mathbf{\boldsymbol\theta})} = {\rm e}^{-(\tau_{\rm gas} + \tau_d)}.
\end{equation}
If we assume that dust is completely sublimated inside \HII regions, we find that the ratio of the dust extinction optical depth to the gas optical depth, $\tau_{d} / \tau_{gas}$, is below $ \num{8e-4} $ along any line of sight. This is estimated by taking the peak value of the fitting formula for $\tau_{d}(\nu)$, that is $\tau_{d} \approx 5 \ N_{H} / (10^{21} ~{\rm cm}^{-2})$. In this scenario the effect of dust is always negligible in our simulations.
Estimates based on observations and numerical simulations \citep{Inoue2002,Ishiki2018}, have shown that radiation pressure creates a dust cavity inside \HII regions, with a typical size of $\sim 30\%$ of Str\"omgren radius. It has also been shown that the grain size distribution is less affected by the radiation from a star cluster than by a single O or B star.

In this section, we estimate the effects of dust extinction on \fesc by assuming no sublimation, therefore setting an upper limit on the effect of dust. In this case, the dust column density is directly proportional to the total hydrogen column density:
\begin{equation}
    \tau_{\rm d}(\nu) = N_{\rm H}\left(\frac{Z}{Z_0}\right) \sigma_{\rm d,eff}(\nu),
\end{equation}
where $Z_0 = 0.2 Z_\odot$ is the gas-phase metallicity of the SMC and we use the fitting formula from \cite{Gnedin2008} for the effective cross section $\sigma_{\rm d,eff}(\nu)$. 
\begin{figure*}
    \centering
    \includegraphics[width=0.8\textwidth]{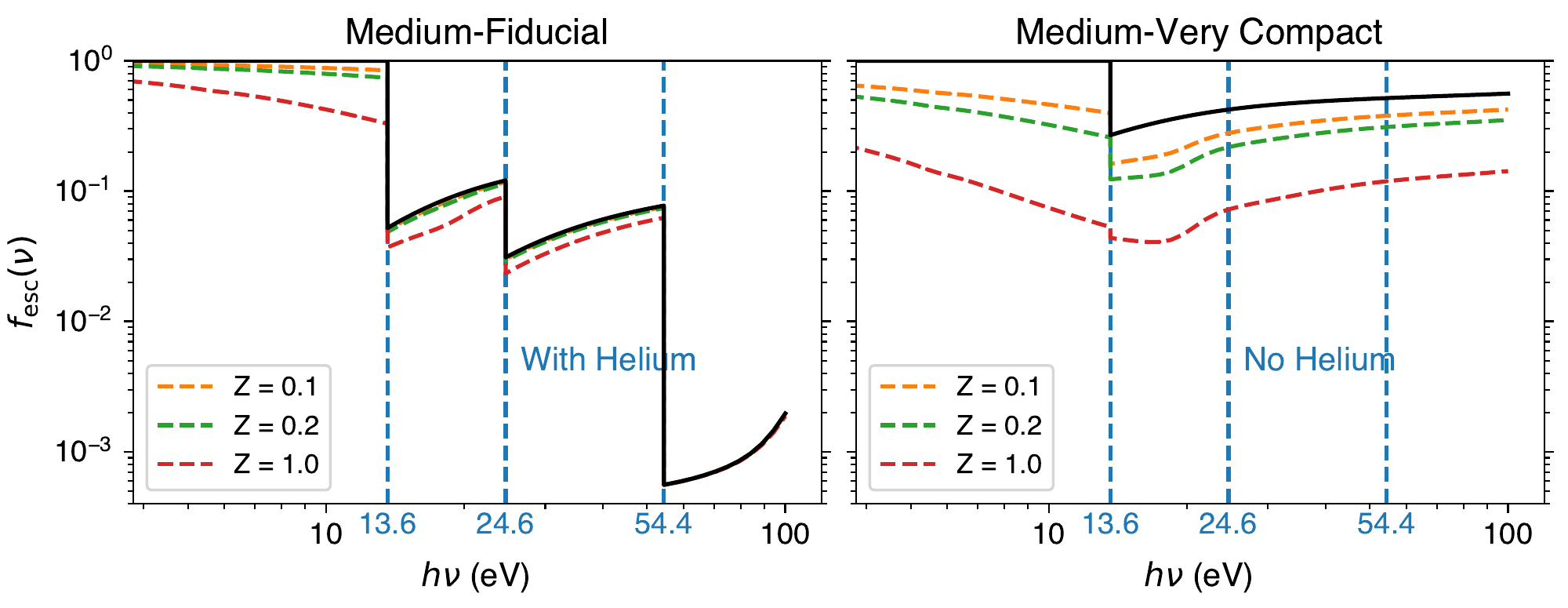}
    \caption{Escape fraction of photons as a function of $h \nu$ from two of the clouds: Medium-Fiducial (left) and Medium-Very Compact (right). From the simulation with He and He$^+$ ionizing photons enabled, we observe that the escape fraction of He ionizing photons is nearly identical to the escape fraction of H ionizing photons. Stars generally do not emit enough high energy photons to ionize He$^+$, hence $f_{\rm esc}(\nu)$ at the He$^+$-ionizing edge is close to zero.
    }
    \label{fig:fesc_ev}
\end{figure*}

In Figure~\ref{fig:fesc_ev}, we plot the escape fraction, $\langle f_{esc} (\nu) \rangle$, as a function of photon energy. Here $\langle f_{esc} (\nu) \rangle$ is averaged over the whole sky, weighted by the ionising luminosity of stars in the correspondent bin, and averaged over time.
The luminosity per frequency below the hydrogen ionization edge is approximated as a constant fraction of $Q_{H}$, i.e. $L / ({\rm ergs \ s}^{-1}) = c_1 Q_{H} / ({\rm s}^{-1})$, where $c_1$ is constant as a function of stellar mass. As shown in Table~\ref{tab:dust}, we find that dust extinction becomes increasingly dominant with increasing cloud mass and cloud compactness, especially for clouds with $Z = 1.0$~Z$_\odot$. More compact clouds have higher total hydrogen column density, thus higher dust column density, even though \fesc due to dust free gas is large because the neutral hydrogen column density becomes low. The most compact and most massive cloud in the table have $80\%$ reduction of \fesc for gas with solar metallicity, while the reduction is between $3\%$ to $50\%$ for less massive and less compact clouds. The effect of dust on \fesc, however, becomes small or negligible for a gas with metallicity below $1/10$ solar.
\begin{table*}
 \centering
 \caption{
    Escape fraction (percentages) at the Lyman edge with and without dust extinction. We consider four models in the calculation of photon optical depth: pure hydrogen and helium gas and gas plus dust with metallicities $Z = 0.1$, $0.2$, and $1.0$. The numbers highlighted in bold face mark the metallicity at which including dust extinction causes a relative decrease $>20 \%$ with respect to \fesc without dust.
 }
 \label{tab:dust}
 \renewcommand{\arraystretch}{1.3}
 \begin{tabular}{cccccc}
   \toprule
   Compactness
   & Job Names
   & {\fesc}
   & {\fesc$^{+{\rm dust}}_{Z = 0.1}$}
   & {\fesc$^{+{\rm dust}}_{Z = 0.2}$}
   & {\fesc$^{+{\rm dust}}_{Z = 1.0}$} \\
   \hline
Fiducial      & XS-F     & \gray{43.7} $^{\rm a}$ & \gray{43.0} & \gray{42.3} & \gray{37.1} \\
Fiducial      & S-F      & \gray{53.3} & \gray{52.3} & \gray{51.4} & \gray{44.7} \\
Fiducial      & M-F      &  5.2 &  5.0 &  4.9 & {\bf 3.7} \\
Fiducial      & L-F      &  1.3 &  1.2 & 1.1 &  {\bf 0.6} \\
Fiducial      & XL-F     &  0.5 & {\bf 0.4} &  {\bf 0.3} &  {\bf 0.1} \\
Compact       & XS-C     & \gray{91.6} & \gray{91.3} & \gray{91.0} & \gray{88.7} \\
Compact       & S-C      & 23.5 & 22.5 & 21.5 & {\bf 15.1} \\
Compact       & M-C      & 15.8 & 14.5 & 13.3 &  {\bf 7.4} \\
Compact       & L-C      & 13.7 & 11.8 & {\bf 10.3} &  {\bf 3.9} \\
Compact       & L-C-lm   & 35.2 & 31.8 & 28.9 & {\bf 15.5} \\
Very Compact  & XXS-VC   & 78.6 & 77.6 & 76.5 & 68.9 \\
Very Compact  & XS-VC    & 63.2 & 59.9 & 56.8 & {\bf 37.9} \\
Very Compact  & S-VC     & 39.7 & 35.5 & {\bf 31.8} & {\bf 14.5} \\
Very Compact  & M-VC     & 26.9 & {\bf 16.2} & {\bf 12.3} &  {\bf 4.4} \\
   \hline
   \multicolumn{6}{p{10cm}}{$^{\rm a}$ The gray data in this table is from the `XS-F', `S-F', and `XS-C' clouds where the simulation results are less reliable because the SFE is overestimated due to missing feedback processes in low-mass stars (see Paper~I).}
 \end{tabular}
\end{table*}

\section{Discussion}
\label{sec:disc}

\subsection{Analytic Modelling and Interpretation of \texorpdfstring{\fesc{}}{fesc}}
\label{sec:model}

In this section we investigate the trends observed in the simulation for \fesc, using a simple analytic model to better understand the dominant physical processes which determine \fesc, and make informed guesses on the extrapolation of the results to a broader parameter space. 
In this model we ignore dust extinction.

The qualitative trends for \fesc as a function of compactness and cloud mass can be explained rather simply in terms of two timescale: $t_{uv}$ that is the time interval during which the bulk of ionizing radiation is emitted, and $t_{esc}$ that is the typical timescale over which $f_{esc}$ increases from being negligible to unity, that is related to the timescale of the duration of the star formation episode, $t_{SF}$, because UV feedback is responsible for stopping star formation and clearing our the gas in the star cluster. When $t_{esc} \gg t_{uv}$, most of the ionising radiation is absorbed in the cloud and \fesc is very small. In Paper~I we found that $t_{SF}\approx 6 \ t_{cr}$, where
\begin{equation}\label{eq:tcrit}
    t_{cr} = 0.40~{\rm Myr} \left(\frac{m_{gas}}{10^4~M_\odot}\right)^{1/3} \left(\frac{\overline n_{gas}}{10^3~cm^{-3}}\right)^{-1/3},
\end{equation}
is the sound crossing time (assuming $c_s = 10$~km/s), which increases with the mass of the cloud and decreases with increasing compactness of the cloud. 

In other words, \fesc in the two most massive fiducial clouds is very small because massive stars are short lived with respect to the star formation timescale of the cloud, therefore they spend most of their life on the main sequence deeply embedded inside the gas rich molecular cloud and their radiation is mostly absorbed. Vice versa, the very compact clouds form all their stars and expel/consume their gas on a timescale shorter than $t_{uv} \sim 3$~Myr, therefore \fesc is closer to unity.

Next we describe the quantitative details of our analytic model for \fesc, that we will show can reproduce quite accurately the simulation results. Informed by the results of the simulations, we assume that $f_{esc}(t)$ grows linearly with time from a value of zero at time $t\le t_{in}$ to a maximum value $f_{esc}^{\rm max}$ at time $t_{esc}$:
\begin{equation}\label{eq:fesc}
f_{esc}(t)=
\begin{cases}
0 &\text{if}~t<t_{in},\\
\frac{t-t_{in}}{t_{esc}} &\text{if}~t_{in} \le t<t_{in}+t_{esc},\\
1 &\text{if}~t\ge t_{in}+t_{esc}.
\end{cases}
\end{equation}
For the sake of simplicity, we model the UV burst as a simple top-hat function with origin at $t=0$ and width $t_{uv}$. This assumption appears to be a good approximation for most of the simulations (see Figure~\ref{fig:Qesc}) because the dominant fraction of the ionising radiation is emitted by the most massive stars in the star cluster that have a rather constant main-sequence lifetime as a function of their mass, $t_{MS} \sim 3$~Myr, for masses above $\sim 30\msun$. 
The mass of the most massive star in the cluster, $M_{*,max}$, correlates with the mass of the star cluster, $m_{cl}$, according to the relationship (see Paper~I):
\begin{equation}\label{eq:mmax}
    M_{max} \approx 205~M_\odot \left(\frac{m_{cl}}{10^4 \msun}\right)^{0.66}.
\end{equation}
Note that Eq.~(\ref{eq:mmax}) is a numerical fit to the simulation data, and it seems to overestimate M$_{max}$ in massive clouds, likely due to our finite resolution and the inability to fully resolve sink fragmentation.
We then convert this mass to the main-sequence lifetime using Eq.~(\ref{eq:tstar}) and set $t_{uv}=t_{MS}(M_{max})$.

Our assumption may fail for the cases in which \fesc is very small (the most massive fiducial clouds), because $f_{esc}(t)$ remains negligibly small for nearly the duration of the life on the main sequence of massive stars (\ie, $t_{uv} \simlt t_{in}$), and only slightly less massive stars are able to stay on the main sequence long enough when $f_{esc}(t)$ starts to rise to larger values.

\begin{figure}
    \centering
    \includegraphics[width=\columnwidth]{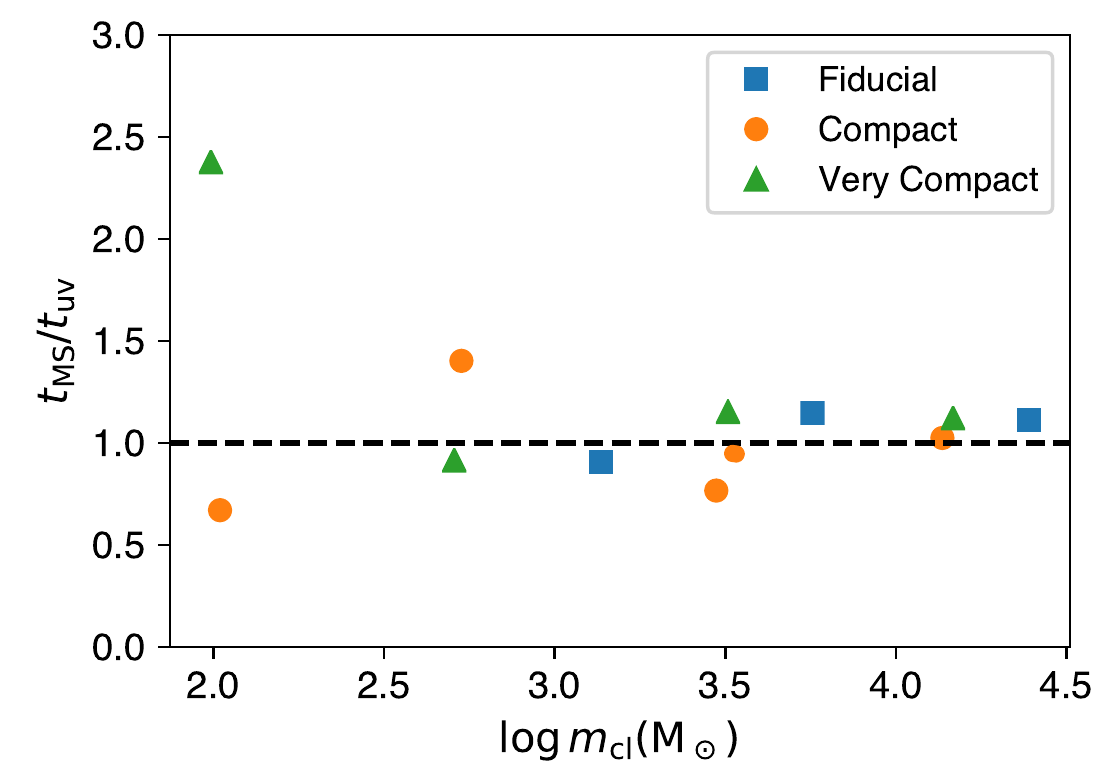}
    \caption{Ratio of $t_{MS}({M_{max})}$ to the measured $t_{uv}$. The $t_{uv}$ is measured as the  Full-Width Half-Maximum of the $Q(t)$ curve.}
    \label{fig:com_tuv}
\end{figure}

In order to test this assumption we compare $t_{MS}$ calculated as explained above, with the values of $t_{uv}$ measured in the simulations as the full-width half maximum of the $Q_{esc}(t)$ curve. Figure~\ref{fig:com_tuv} shows that indeed $t_{MS}(M_{max})/t_{uv}$ is close to unity with small scatters, even for the fiducial clouds, demonstrating the goodness of our assumption. 

With these two simple assumptions on the shape of $f_{esc}(t)$ and $Q(t)$, we find that the time-averaged \fesc is:
\begin{equation}\label{eq:mfesc}
\fesc=
\begin{cases}
\frac{t_{uv}-t_{in}}{t_{uv}} - \frac{1}{2}\frac{t_{esc}}{t_{uv}} &\text{if}~t_{esc}< (t_{uv} - t_{in}),\\
\frac{1}{2}\frac{(t_{uv}-t_{in})^2}{t_{uv} t_{esc}} &\text{if} ~t_{esc}\ge (t_{uv} - t_{in}).
\end{cases}
\end{equation}

\begin{figure}
    \centering
    \includegraphics[width=.9\columnwidth]{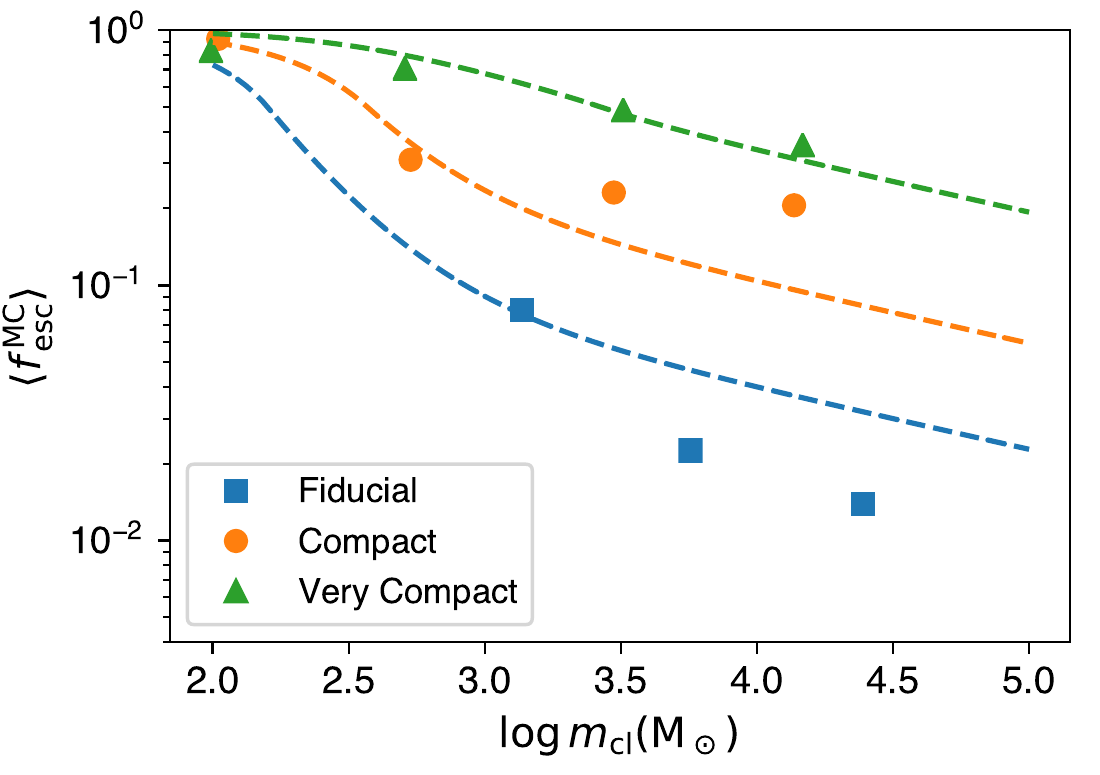}
    \includegraphics[width=.9\columnwidth]{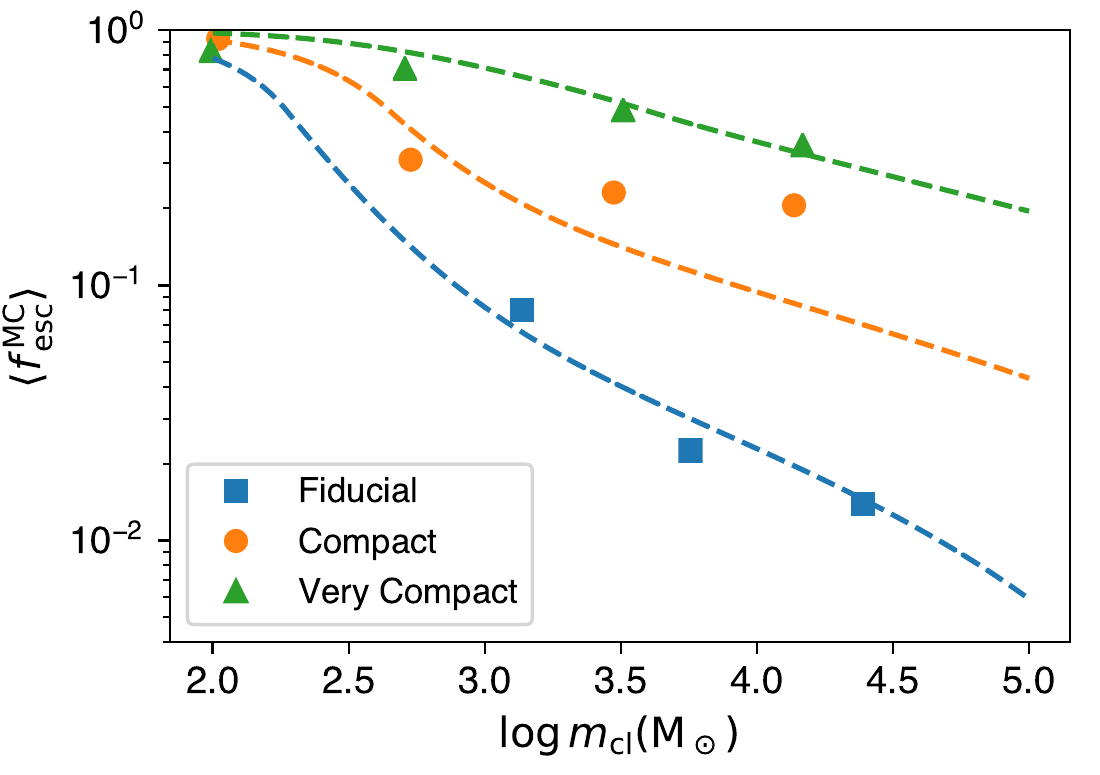}
    \caption{
    Comparing model \fesc (dashed lines) with \fesc from simulations (shapes). The models have $t_{esc}$ (top) or $t_{in}$ and $t_{esc}$ (bottom) as parameters. Both models work equally well on the Compact and Very Compact clouds while only the latter model works well on the Fiducial clouds. 
    Bottom: The modeled \fesc using pure cloud parameters. Eq.~(\ref{eq:mfesc}) and (\ref{eq:tcr}) are used.}
    \label{fig:compareR}
\end{figure}
\hide{
\begin{figure}
    \centering
    \includegraphics[width=.9\columnwidth]{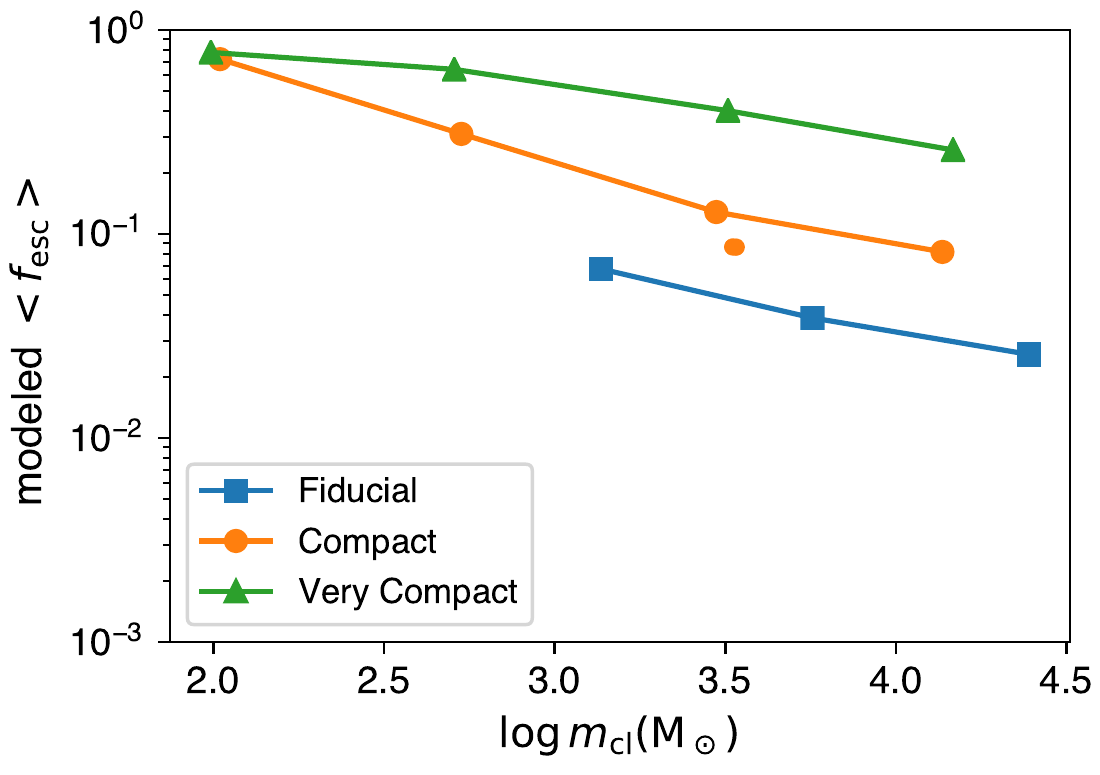}
    \includegraphics[width=.9\columnwidth]{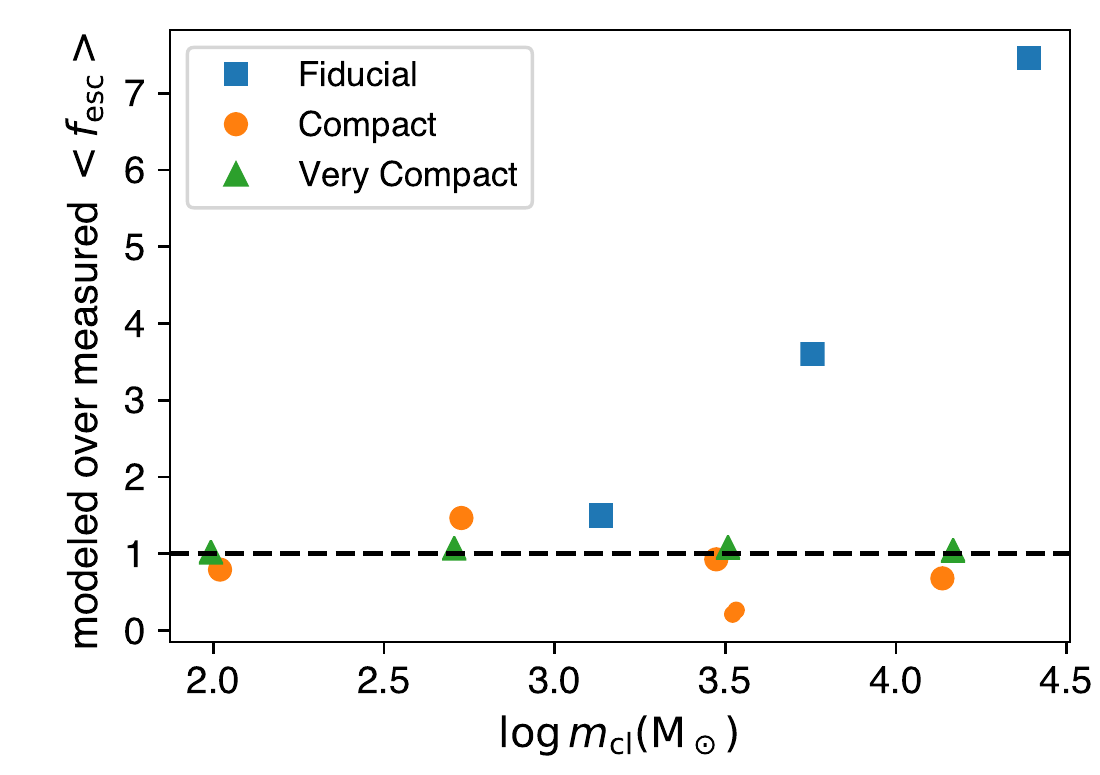}
    \caption{
    Top: The modeled escape fraction of ionizing photons \fesc = $S_{\rm esc}/S$. 
    Bottom: The ratio of the modeled to the measured \fesc.
    }
    \label{fig:R_param}
\end{figure}
}

\begin{figure}
    \centering
    \includegraphics[width=.9\columnwidth]{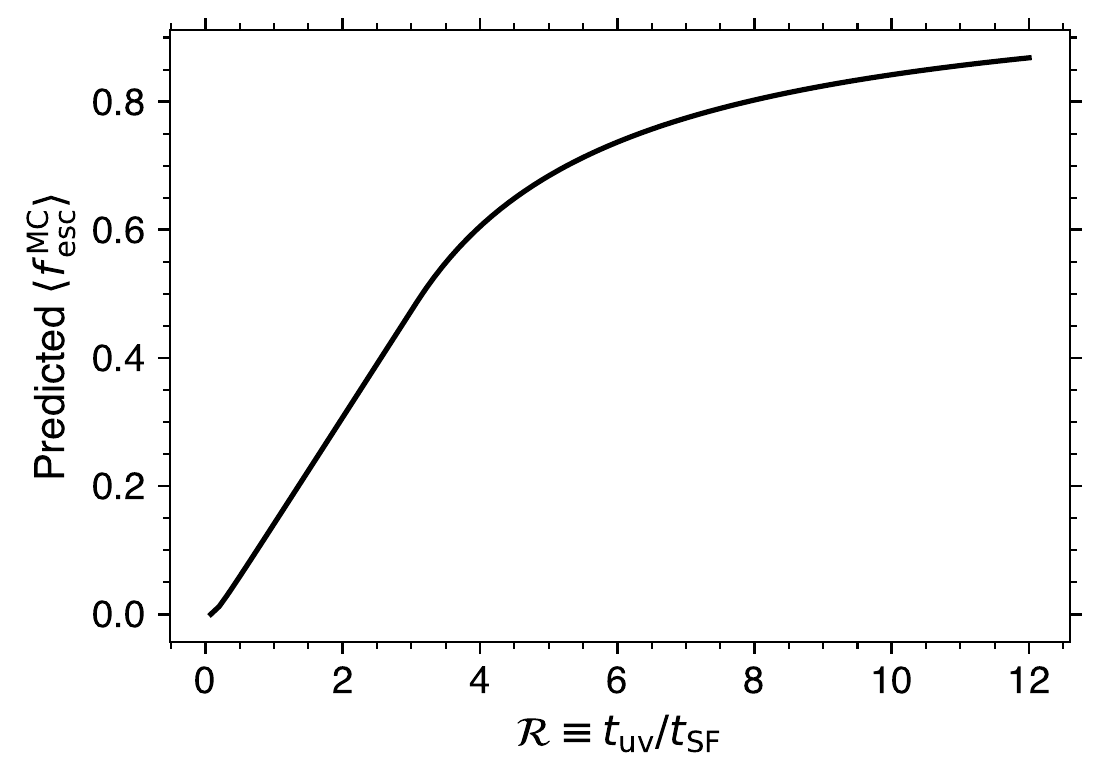}
    \caption{Conversion from the ${\cal R}$ parameter to \fesc, following Eq.~(\ref{eq:mfesc2}).}
    \label{fig:newR}
\end{figure}

Guided by a physically motivated prior for $t_{esc}$ and $t_{in}$, we found that they are both proportional to $t_{SF} \propto t_{cr}$, being the timescale over which feedback is able to destroy the molecular cloud and stop star formation. 

Assuming $t_{uv} = t_{MS}(M_{max})$, we fit Eq.~(\ref{eq:mfesc}) to the data, using $t_{in}/t_{cr}$ and $t_{esc}/t_{cr}$ as free parameters. In Figure~\ref{fig:compareR} we show the best fits compared to the data for two models: in the top panel we fit the data with a one-parameter model by  setting $t_{in}=0$ (hence \fesc $= 1 - 0.5 t_{esc} / t_{uv}$ when $t_{esc} < t_{uv}$ and $0.5 t_{uv} / t_{esc}$ otherwise). The best fit parameter in this model is $t_{esc} = 21 t_{cr} \approx 3.5 t_{SF}$, where we have used $t_{SF}=6 t_{cr}$, found for simulations with gas at solar metallicity (see Paper~I).
This model works well for the Very Compact clouds and slightly underestimates \fesc for massive Compact clouds by a factor of $\lesssim 2$. It also overestimates \fesc for the Fiducial clouds where the lifetime of the most massive star ($\sim 3$~Myr) is shorter than several free-fall times and UV radiation is shut down before the gas is expelled, resulting in \fesc below $10\%$.

The bottom panel of Figure~\ref{fig:compareR} shows the two-parameter model in Eq.~(\ref{eq:mfesc}). This model resolves the discrepancy between the model-predicted \fesc and the simulation results from the massive fiducial clouds. This model, similar to the one-parameter model, slightly underestimates \fesc from the massive Compact clouds. We believe that part of the discrepancy is due to second order effects from weighting \fesc over the stellar spectra of different mass stars. As shown in Table~\ref{tab:2}, \fesc at the Lyman edge from these clouds, being significantly smaller, is closer to the model predictions. For this model the best fit parameters are $t_{in} = 0.5 t_{cr} \approx 0.08 t_{SF}$ and $t_{esc} = 18 t_{cr} \approx 3 t_{SF}$. In both models we find that at the end of the star formation episode (at $t=t_{SF}$) the value of the escape fraction is $f_{esc}(t=t_{SF}) \sim 30\%$ (see Eq.~(\ref{eq:fesc})), and this value keeps increasing approximately linearly as a function of time after that.

Hence, if we define ${\cal R}\equiv t_{uv}/t_{SF}$, using the best fit parameters for the two-parameters model, we can rewrite Eq.~(\ref{eq:mfesc}) as 
\begin{equation}\label{eq:mfesc2}
\fesc =
\begin{cases}
1- \frac{1.58}{\cal R} &\text{if} \ {\cal R}>3.1, 
\\
0.167 \ \frac{({\cal R} - 0.08)^2}{{\cal R}} 
&\text{if} \ 0.08\le {\cal R} \le 3.1.
\end{cases}
\end{equation}
Eq.~(\ref{eq:mfesc2}) is shown in Figure~\ref{fig:newR}. 
Due to the non-linear term $({\cal R} - 0.08)^2 / {\cal R}$, when ${\cal R} \lesssim 1$, \fesc becomes very small and approaches zero as ${\cal R} \rightarrow 0.08$. This is the limit when $t_{uv}=t_{in}$ and all massive stars have died by the time $f_{esc}(t)>0$. In this limit our model assumption fails and we need to consider longer lived (less massive) stars. But for these cases we expect \fesc $\ll 1\%$. 
When ${\cal R} \lesssim 3$ (or \fesc$<50\%$), \fesc is roughly proportional to ${\cal R}$:  \fesc$\sim 0.17 {\cal R}$.

This equation can help us interpret the results on \fesc for simulations with gas at sub-solar metallicity. In Paper~I we found that for gas metallicitities $<1/10$~Z$_\odot$, the duration of the star formation in the cloud was reduced by roughtly 1/2 (\ie, $t_{SF}=3t_{cr}$). 
Hence, for a given molecular cloud mass and compactness, we expect that ${\cal R}$ is roughly twice the value found for solar metallicity, and \fesc is also roughly twice as large if \fesc$<50\%$. We also note that lowering the metallicity reduces the SFE of the cloud, hence for a given molecular cloud mass, the mass of the star cluster is reduced and \fesc increases with respect to the solar metallicity case. The overall effect is a strong sensitivity of \fesc on the gas metallicity for two clusters of equal stellar mass.

Using the results in Paper~I for a cloud at solar metallicity we can write ${\cal R}$ as a function of the cloud's parameters. For star masses $M>10$~M$_\odot$ we can approximate $t_{uv} = t_{MS}=2.86+1.9\times10^3 (M/\msun)^{-2}$~Myr and using Eq.~(\ref{eq:mmax}) we have 
\begin{equation}\label{eq:tuv}
t_{uv}=2.86+0.045 m_{cl,4}^{-1.32} ~{\rm Myr}
\end{equation}
where $m_{cl,4} \equiv m_{cl}/10^{4}~M_\odot$,
For clouds with solar metallicity, we can also write $t_{cr}$ in Eq.~(\ref{eq:tcrit}) as a function of $m_{cl}$ and the cloud compactness, by expressing $m_{gas}$ as a function of the cluster mass using the following relationship found in Paper~I (valid for clouds at solar metallicity):
\begin{equation}
  m_{\rm cl} = 200~{\rm M}_{\odot} \cdot \left(\frac{m_{gas}}{10^4\msun}\right)^{1.4} \left( 1+\frac{\overline{n}_{gas}}{n_{\rm cri}}\right)^{0.91} + m_{fl} \, ,
  \label{eq:mcl1}
\end{equation}
where $n_{\rm cri} \approx 10^3 ~\pcc$ is the critical density and $m_{fl}=10$~M$_\odot$ is the mass floor. Therefore, neglecting the mass floor (\ie, $m_{fl}=0$), since $t_{SF}=6t_{cr}$, we find:
\begin{equation}\label{eq:tcr}
    {\cal R} =
    (0.473+0.008 m_{cl,4}^{-1.32}) \ m_{cl,4}^{-0.24} \left(\frac{\overline n_{gas}}{n_{cri}}\right)^{0.33} \left(1+\frac{\overline n_{gas}}{n_{cri}}\right)^{0.22}.
\end{equation}

\subsection{Ionising Photons from OB Associations}\label{sec:ob}

In our Galaxy and nearby dwarf and spiral galaxies, the mass function of young massive star clusters (or OB associations) is a power-law with slope 
$\xi \simeq -2 \pm 0.5$ \citep{Rosolowsky:2005,Hopkins:2012b}:
\[
\frac{dN}{dm_{\rm cl}} = A m_{\rm cl}^{\xi},
\]
where, assuming $\xi=-2$ \citep{Hopkins:2012b}, we find $A=M_{*,gal}/\Lambda$, with $\Lambda= \ln{(m_{cl}^{\rm max}/m_{cl}^{\rm min}})$. Assuming $m_{cl}^{\rm max}=10^6$~M$_\odot$ and $m_{cl}^{\rm min}=100$~M$_\odot$, we estimate $\Lambda \approx 9.2$.
Therefore, assuming an escape fraction \fesci{} from the atomic phase of the ISM in the galaxy (defined excluding the absorption due to the molecular cloud) that is constant as a function of the cluster mass, we find:

\begin{align}
    \begin{split}
    S_{esc}^{\rm gal}&=\fesci \int_{m_{cl}^{\rm min}}^{m_{cl}^{\rm max}} \frac{\md N}{\md m_{cl}} S_{esc}(m_{cl}) \ {\rm d} m_{cl}\\
    &= \fesci \frac{M_{*,\rm gal}}{\Lambda} \int_{m_{cl}^{\rm min}}^{m_{cl}^{\rm max}} m_{cl}^{-1} \frac{S_{esc}}{m_{cl}} \ {\rm d} m_{cl}\\
    &\approx \fesci \left(\frac{M_{*,gal}}{1\msun}\right) \cdot
    \begin{cases}
    \frac{\num{7.4e60}}{\Lambda/9.2}
    \left(\frac{m_{cl}^{\rm max}}{10^6~M_\odot}\right)^{0.4}~&\text{(very compact}),\\
    \num{1.4e60}~&\text{(compact)}),\\
    \num{1.8e59}~&\text{(fiducial}).
    \end{cases}
    \end{split}
\end{align}
Therefore, as anticipated before in Section~\ref{sec:ef}, in the local Universe (fiducial clouds) the escaping ionising radiation from a galaxy is produced by roughly equal contribution from small and large mass star clusters, and the number of escaping photons is $\sim 10^{59}$ per unit solar mass in stars. Therefore, the total escaping radiation is quite insensitive to the upper and lower mass limits of the mass distribution of OB associations. Compact star clusters are similar but with $\sim 10$ times more ionizing photons per mass in stars.
For very compact clouds (100 times denser than the fiducial clouds) the escaping ionising radiation is dominated by the few most massive star clusters in the galaxy, and the number of escaping photons per units star mass is about 40 times higher than for the fiducial clouds. 

Also, if we make the simple assumption that the mass of the most massive star cluster is related to the total stellar mass $M_{*,gal}$ of the galaxy, by setting $\int_{m^{\rm max}_{cl}}^{+\infty} dN/d\ln{m_{cl}}=1$, we find $m_{cl}^{\rm max} \sim M_{*,gal}/\Lambda$. Hence, if star clusters in high-redshift galaxies form in very compact molecular clouds, massive galaxies would be more efficient contributor to propel reionization than dwarf galaxies. Of course the discussion above is only valid if \fesci is constant not only as a function of the star cluster mass but also as a function of the mass of the galaxy.

Similarly to $S_{esc,tot}$, we can estimate the total emitted ionising radiation by OB association:
\begin{align}
    S_{tot}&=\int \frac{dN}{dm_{cl}} S(m_{cl}) \mathop{d m_{cl}}\\
    &\approx
      \frac{\num{1.2e62}}{\Lambda}\left(\frac{M_{*,gal}}{1~M_\odot}\right)\left(\frac{m_{cl}^{\rm
      max}}{10^6~M_\odot}\right)^{0.4},
\end{align}
and the mean escape fraction from a galaxy by taking the ratio $S_{esc}^{\rm gal}/S_{tot}$:
\begin{align}\label{eq:fgal}
\begin{split}
\fescg \approx \fesci \cdot
\begin{cases}
56.7\%~&\text{(Very Compact)},\\
10.7\%\left(\frac{\Lambda}{9.2}\right)\left(\frac{m_{cl}^{\rm max}}{10^6~M_\odot}\right)^{-0.4}~&\text{(Compact)},\\
1.4\%\left(\frac{\Lambda}{9.2}\right)\left(\frac{m_{cl}^{\rm max}}{10^6~M_\odot}\right)^{-0.4}~&\text{(Fiducial)}.
\end{cases}
\end{split}
\end{align}
This last equation confirms that \fescg from galaxies in the local Universe (fiducial clouds) is extremely small \fescg $\approx \fesci \times 1.4\%$, and only assuming that molecular clouds at redshift $z>6$ were 100$\times$ denser than in the local Universe is possible to propel reionization with UV radiation from massive stars in galaxies.

\section{Summary and Conclusions}
\label{sec:summary}
 
In this paper, the second of a series, we calculate the hydrogen and helium ionizing radiation escaping realistic young star cluster forming in turbulent molecular clouds. To the best of our knowledge this is the first work in which \fesc is calculated by self-consistently simulating the formation, UV radiation feedback, and contribution to the escaping ionising radiation from individual massive stars producing the observed IMF slope and normalization. We used a set of high-resolution radiation-magneto-hydrodynamic simulations of star formation in self-gravitating, turbulent molecular clouds presented in He, Ricotti and Geen (2019), in which we vary 
the mass of the star forming molecular clouds between $m_{\rm gas}=10^3$~M$_\odot$ to $3 \times 10^5$~M$_\odot$ and adopt gas densities typical of clouds in the local universe ($\overline n_{\rm gas} \sim 1.8\times 10^2$~cm$^{-3}$), and 10$\times$ and 100$\times$ denser, expected to exist in high-redshift galaxies.

We find that \fesc decreases with increasing mass of the star cluster and with decreasing initial gas density. Molecular clouds with densities typically found in the local Universe have negligible \fesc, ranging between $8\%$ to $1.4\%$ for clouds with masses ranging from $\num{3e4}$ to $\num{3e5} \msun$.
Ten times denser molecular clouds have \fesc$\approx 20\%-30\%$, while $100\times$ denser clouds, which produce globular cluster progenitors, have \fesc$\approx 30\%-50\%$. 
Star clusters with mass $\lesssim 500$~M$_\odot$ have $\fesc > 50\%$ independently of their compactness but assuming the observed OB association luminosity function, $dN/dm_{cl} \propto m_{cl}^{-2}$, fall short in providing the required ionising photons for reionization.

We reproduce the simulation results for \fesc using a simple analytic model, in which the observed trends with cloud mass and density are understood in terms of the parameter ${\cal R}$, the ratio of the lifetime of the most massive star in the cluster to the star formation timescale, that, for clouds with solar metallicity is about 6 times the sound crossing time of the cloud. We find that it takes about 20 times the sound-crossing time ($t_{\rm cr} = r_{\rm gas}/10~$km/s), or 3.5$\times$ the star-formation time, for the stars to ionize the cloud and for $f_{\rm esc}(t)$ to become of order of unity. Since $r_{\rm gas}$, therefore $t_{\rm cr}$, increases with increasing cloud mass and decreasing density and
the lifetime of the dominating LyC sources is constant at $\sim 3$~Myr, our model quantitatively reproduce the increase of \fesc with decreasing cloud mass and increasing cloud density, observed in the simulations.

We find that \fesc increases with decreasing gas metallicity, even when ignoring dust extinction, due to stronger LyC radiation feedback and faster ionization of the cloud. However, as the metallicity decreases, the SFE declines, therefore the total number of escaped LyC photons decreases. For the L-C cloud which we use to investigate this effect, the value of $Q_{\rm esc}$ decreases by a factor of 2 as we decrease the metallicity from $Z_{\odot}$ to $0.1 Z_{\odot}$, although the value of \fesc doubles.

We find that in all our simulations the values of \fesc for He LyC photons are nearly identical to \fesc for H LyC photons. We explain this result by  noting that the ionization fronts of \ion{H}{ii} and \ion{He}{ii} are comparable around the dominant sources of ionization, namely hot O stars.

When dust extinction is considered, assuming no sublimation inside \ion{H}{ii} region, \fesc is nearly unaffected compared to dust-free estimates for values of the metallicity $<0.1$ solar (see Table~\ref{tab:dust}). Assuming solar metallicity, while \fesc for the least massive and least compact clouds is nearly unchanged, \fesc for the more massive and more compact clouds is reduced significantly, by up to 
$80\%$. 
SN explosions have little effect on the time-averaged \fesc for nearly all the star clusters considered in this work, unless we consider fiducial clouds (local Universe) with mass $\simgt \num{e5} \msun$. In these simulations SN explosions occur before $f_{\rm esc}(t)$ becomes significantly larger than zero, hence mechanical feedback may increase \fesc.

In conclusion, we find an upper limit on \fescg $<3\%-10\%$ for star clusters forming in molecular clouds similar in compactness to today's clouds (see discussion in \S~\ref{sec:disc} and Eq.~(\ref{eq:fgal})). Therefore, since large scale simulations show that cosmic re-ionization requires $\fescg \simgt 10\%-20\%$, we conclude that the sources of reionization at $z>6$ must have been very compact star clusters forming in molecular clouds about $10$ to $100\times$ denser than in today's Universe. This result indirectly suggests a significant formation of old globular clusters progenitors at redshifts $z>6$.

\section*{ACKNOWLEDGEMENTS}
MR acknowledges the support by NASA grant 80NSSC18K0527. 
The authors acknowledge the University of Maryland supercomputing resources (http://hpcc.umd.edu) made available for conducting the research reported in this paper.

This work has been funded by the European Research Council under the European Community's Seventh Framework Programme (FP7/2007-2013). SG has received funding from Grant Agreement no. 339177 (STARLIGHT) of this programme. SG acknowledges support from a NOVA grant for the theory of massive star formation.

\appendix

\section{Converting column density to escape fraction}
\label{sec:app1}
\begin{figure}
  \centering
  \includegraphics[width=.8\columnwidth]{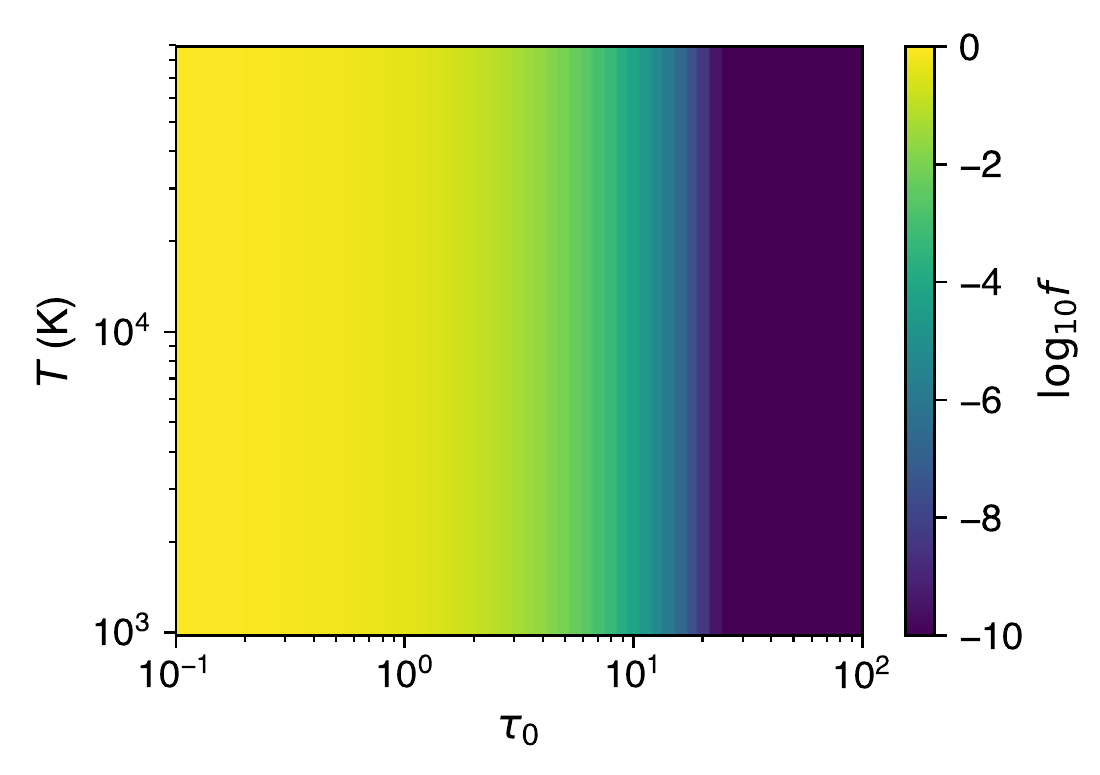}
  \includegraphics[width=.8\columnwidth]{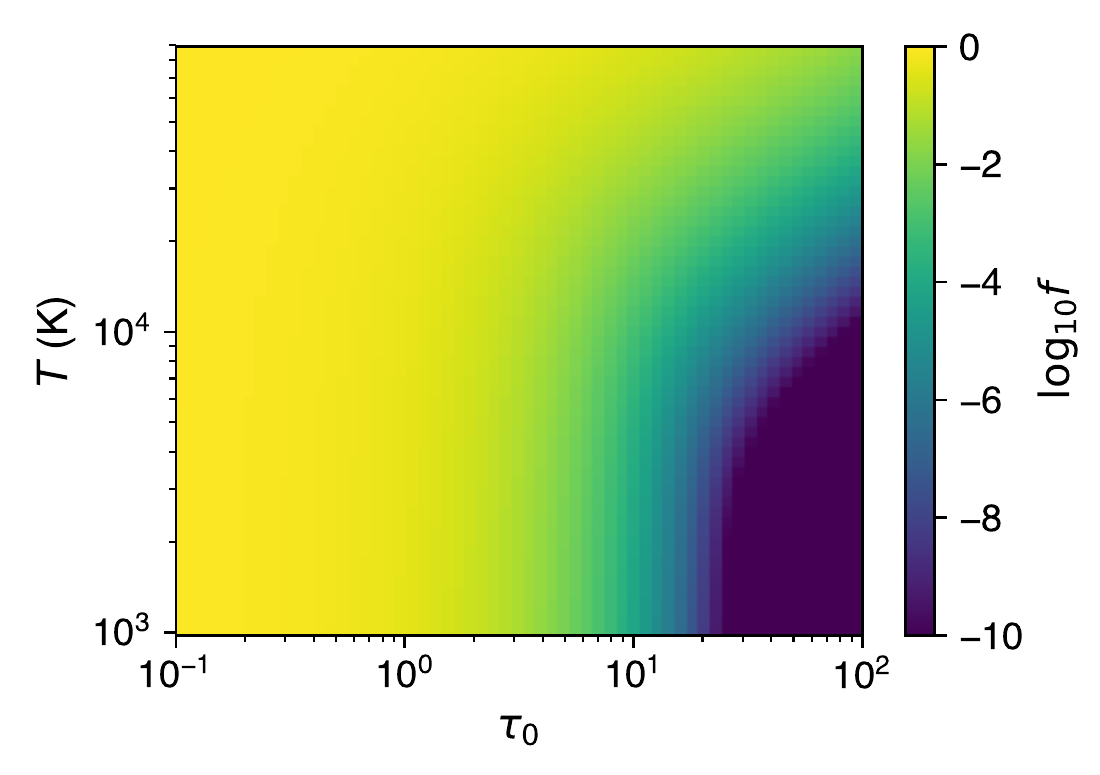}
  \caption{Color plots of $ f(\tau_0, T) $ where $\tau_0 \equiv N_{HI} \sigma_0 / m_p$. Top: Eq.~(\ref{eq:fprox}) assuming $\tau_\nu = \tau_0$;
  Right: Eq.~(\ref{eq:ftau}). At high temperatures,  the escape fraction calculated from Eq.~(\ref{eq:ftau}) is much higher than $ \exp(-N\sigma_0) $ when $ N\sigma_0 > 2 $. This modulation makes the calculated escape fraction  higher than estimated from $ \exp(-N\sigma_0) $.}
  \label{fig:tau}
\end{figure}

A comparison between Eq.~(\ref{eq:ftau}) and Eq.~(\ref{eq:fprox}) is shown in Figure~\ref{fig:tau}. The $x$ axis is $\tau_0 \equiv N_{HI} \sigma_0 / m_p$ and $y$ axis is the surface temperature of a star. On the top panel is $f(\tau_0, T) = \exp (-\tau_0)$. On the bottom panel is $f(\tau_0, T) = f_{esc}(N_{HI}, T)$, following Eq.~(\ref{eq:ftau}). Clearly Eq.~(\ref{eq:ftau}) drops much slower with $ \tau $ than Eq.~(\ref{eq:fprox}) does at high temperatures.
In order to compute Eq.~(\ref{eq:ftau}) effectively, we do an interpolation of it and apply it in our code. 

In the calculation of escape fraction, some classical mass-luminosity 
\citep{Bressan:1993} and mass-radius \\
\citep{Demircan:1991} relations are used.

\bibliographystyle{mnras}
\bibliography{HRG19b.bib}

\begin{thebibliography}{}
\makeatletter
\relax
\def\mn@urlcharsother{\let\do\@makeother \do\$\do\&\do\#\do\^\do\_\do\%\do\~}
\def\mn@doi{\begingroup\mn@urlcharsother \@ifnextchar [ {\mn@doi@}
  {\mn@doi@[]}}
\def\mn@doi@[#1]#2{\def\@tempa{#1}\ifx\@tempa\@empty \href
  {http://dx.doi.org/#2} {doi:#2}\else \href {http://dx.doi.org/#2} {#1}\fi
  \endgroup}
\def\mn@eprint#1#2{\mn@eprint@#1:#2::\@nil}
\def\mn@eprint@arXiv#1{\href {http://arxiv.org/abs/#1} {{\tt arXiv:#1}}}
\def\mn@eprint@dblp#1{\href {http://dblp.uni-trier.de/rec/bibtex/#1.xml}
  {dblp:#1}}
\def\mn@eprint@#1:#2:#3:#4\@nil{\def\@tempa {#1}\def\@tempb {#2}\def\@tempc
  {#3}\ifx \@tempc \@empty \let \@tempc \@tempb \let \@tempb \@tempa \fi \ifx
  \@tempb \@empty \def\@tempb {arXiv}\fi \@ifundefined
  {mn@eprint@\@tempb}{\@tempb:\@tempc}{\expandafter \expandafter \csname
  mn@eprint@\@tempb\endcsname \expandafter{\@tempc}}}

\bibitem[\protect\citeauthoryear{{Bleuler} \& {Teyssier}}{{Bleuler} \&
  {Teyssier}}{2014}]{Bleuler:2014}
{Bleuler} A.,  {Teyssier} R.,  2014, \mn@doi [\mnras] {10.1093/mnras/stu2005},
  \href {http://adsabs.harvard.edu/abs/2014MNRAS.445.4015B} {445, 4015}

\bibitem[\protect\citeauthoryear{{Bowman}, {Rogers}, {Monsalve}, {Mozdzen}  \&
  {Mahesh}}{{Bowman} et~al.}{2018}]{Bowman:2018}
{Bowman} J.~D.,  {Rogers} A. E.~E.,  {Monsalve} R.~A.,  {Mozdzen} T.~J.,
  {Mahesh} N.,  2018, \mn@doi [\nat] {10.1038/nature25792}, \href
  {https://ui.adsabs.harvard.edu/abs/2018Natur.555...67B} {555, 67}

\bibitem[\protect\citeauthoryear{{Boylan-Kolchin}}{{Boylan-Kolchin}}{2018}]{Boylan-Kolchin:2018}
{Boylan-Kolchin} M.,  2018, \mn@doi [\mnras] {10.1093/mnras/sty1490}, \href
  {https://ui.adsabs.harvard.edu/abs/2018MNRAS.479..332B} {479, 332}

\bibitem[\protect\citeauthoryear{{Bressan}, {Fagotto}, {Bertelli}  \&
  {Chiosi}}{{Bressan} et~al.}{1993}]{Bressan:1993}
{Bressan} A.,  {Fagotto} F.,  {Bertelli} G.,   {Chiosi} C.,  1993, \aaps, \href
  {http://adsabs.harvard.edu/abs/1993A%26AS..100..647B} {100, 647}

\bibitem[\protect\citeauthoryear{{Bridge} et~al.,}{{Bridge}
  et~al.}{2010}]{Bridge2010}
{Bridge} C.~R.,  et~al., 2010, \mn@doi [\apj] {10.1088/0004-637X/720/1/465},
  \href {https://ui.adsabs.harvard.edu/abs/2010ApJ...720..465B} {720, 465}

\bibitem[\protect\citeauthoryear{{Dale}, {Ngoumou}, {Ercolano}  \&
  {Bonnell}}{{Dale} et~al.}{2014}]{Dale:2014}
{Dale} J.~E.,  {Ngoumou} J.,  {Ercolano} B.,   {Bonnell} I.~A.,  2014, \mn@doi
  [\mnras] {10.1093/mnras/stu816}, \href
  {http://adsabs.harvard.edu/abs/2014MNRAS.442..694D} {442, 694}

\bibitem[\protect\citeauthoryear{{Demircan} \& {Kahraman}}{{Demircan} \&
  {Kahraman}}{1991}]{Demircan:1991}
{Demircan} O.,  {Kahraman} G.,  1991, \mn@doi [\apss] {10.1007/BF00639097},
  \href {http://adsabs.harvard.edu/abs/1991Ap%26SS.181..313D} {181, 313}

\bibitem[\protect\citeauthoryear{{Doran} et~al.,}{{Doran}
  et~al.}{2013}]{Doran:2013}
{Doran} E.~I.,  et~al., 2013, \mn@doi [\aap] {10.1051/0004-6361/201321824},
  \href {https://ui.adsabs.harvard.edu/abs/2013A&A...558A.134D} {558, A134}

\bibitem[\protect\citeauthoryear{{Draine}}{{Draine}}{2011}]{Draine:2011}
{Draine} B.~T.,  2011, {Physics of the Interstellar and Intergalactic Medium}.
Princeton University Press

\bibitem[\protect\citeauthoryear{{Ellis} et~al.,}{{Ellis}
  et~al.}{2013}]{Ellis2013}
{Ellis} R.~S.,  et~al., 2013, \mn@doi [\apjl] {10.1088/2041-8205/763/1/L7},
  \href {https://ui.adsabs.harvard.edu/abs/2013ApJ...763L...7E} {763, L7}

\bibitem[\protect\citeauthoryear{Gnedin, Kravtsov  \& Chen}{Gnedin
  et~al.}{2008}]{Gnedin2008}
Gnedin N.~Y.,  Kravtsov A.~V.,   Chen H.-W.,  2008, \mn@doi [\apj]
  {10.1086/524007}, 672, 765

\bibitem[\protect\citeauthoryear{{Hartley} \& {Ricotti}}{{Hartley} \&
  {Ricotti}}{2016}]{Hartley:2016}
{Hartley} B.,  {Ricotti} M.,  2016, \mn@doi [\mnras] {10.1093/mnras/stw1562},
  \href {https://ui.adsabs.harvard.edu/abs/2016MNRAS.462.1164H} {462, 1164}

\bibitem[\protect\citeauthoryear{{He}, {Ricotti}  \& {Geen}}{{He}
  et~al.}{2019}]{He2019}
{He} C.-C.,  {Ricotti} M.,   {Geen} S.,  2019, \mn@doi [\mnras]
  {10.1093/mnras/stz2239}, \href
  {https://ui.adsabs.harvard.edu/abs/2019MNRAS.489.1880H} {489, 1880}

\bibitem[\protect\citeauthoryear{Hopkins}{Hopkins}{2012}]{Hopkins:2012b}
Hopkins P.~F.,  2012, \mn@doi [Monthly Notices of the Royal Astronomical
  Society] {10.1111/j.1365-2966.2012.20730.x}, 423, 2016

\bibitem[\protect\citeauthoryear{{Howard}, {Pudritz}  \& {Harris}}{{Howard}
  et~al.}{2017}]{Howard:2017}
{Howard} C.~S.,  {Pudritz} R.~E.,   {Harris} W.~E.,  2017, \mn@doi [\mnras]
  {10.1093/mnras/stx1363}, \href
  {http://adsabs.harvard.edu/abs/2017MNRAS.470.3346H} {470, 3346}

\bibitem[\protect\citeauthoryear{{Howard}, {Pudritz}, {Harris}  \&
  {Klessen}}{{Howard} et~al.}{2018}]{Howard:2018}
{Howard} C.~S.,  {Pudritz} R.~E.,  {Harris} W.~E.,   {Klessen} R.~S.,  2018,
  \mn@doi [\mnras] {10.1093/mnras/stx3276}, \href
  {http://adsabs.harvard.edu/abs/2018MNRAS.475.3121H} {475, 3121}

\bibitem[\protect\citeauthoryear{{Inoue}}{{Inoue}}{2002}]{Inoue2002}
{Inoue} A.~K.,  2002, \mn@doi [\apj] {10.1086/339788}, \href
  {https://ui.adsabs.harvard.edu/abs/2002ApJ...570..688I} {570, 688}

\bibitem[\protect\citeauthoryear{{Ishiki}, {Okamoto}  \& {Inoue}}{{Ishiki}
  et~al.}{2018}]{Ishiki2018}
{Ishiki} S.,  {Okamoto} T.,   {Inoue} A.~K.,  2018, \mn@doi [\mnras]
  {10.1093/mnras/stx2833}, \href
  {https://ui.adsabs.harvard.edu/abs/2018MNRAS.474.1935I} {474, 1935}

\bibitem[\protect\citeauthoryear{{Izotov}, {Worseck}, {Schaerer}, {Guseva},
  {Thuan}, {Fricke}  \& {Orlitov{\'a}}}{{Izotov} et~al.}{2018}]{Izotov:2018}
{Izotov} Y.~I.,  {Worseck} G.,  {Schaerer} D.,  {Guseva} N.~G.,  {Thuan} T.~X.,
   {Fricke} Verhamme A.,   {Orlitov{\'a}} I.,  2018, \mn@doi [\mnras]
  {10.1093/mnras/sty1378}, \href
  {https://ui.adsabs.harvard.edu/abs/2018MNRAS.478.4851I} {478, 4851}

\bibitem[\protect\citeauthoryear{{Katz} \& {Ricotti}}{{Katz} \&
  {Ricotti}}{2013}]{KatzR:2013}
{Katz} H.,  {Ricotti} M.,  2013, \mn@doi [\mnras] {10.1093/mnras/stt676}, \href
  {http://adsabs.harvard.edu/abs/2013MNRAS.432.3250K} {432, 3250}

\bibitem[\protect\citeauthoryear{{Katz} \& {Ricotti}}{{Katz} \&
  {Ricotti}}{2014}]{KatzR:2014}
{Katz} H.,  {Ricotti} M.,  2014, \mn@doi [\mnras] {10.1093/mnras/stu1489},
  \href {https://ui.adsabs.harvard.edu/\#abs/2014MNRAS.444.2377K} {444, 2377}

\bibitem[\protect\citeauthoryear{{Khaire}, {Srianand}, {Choudhury}  \&
  {Gaikwad}}{{Khaire} et~al.}{2016}]{Khaire:2016}
{Khaire} V.,  {Srianand} R.,  {Choudhury} T.~R.,   {Gaikwad} P.,  2016, \mn@doi
  [\mnras] {10.1093/mnras/stw192}, \href
  {https://ui.adsabs.harvard.edu/abs/2016MNRAS.457.4051K} {457, 4051}

\bibitem[\protect\citeauthoryear{{Kimm}, {Blaizot}, {Garel}, {Michel-Dansac},
  {Katz}, {Rosdahl}, {Verhamme}  \& {Haehnelt}}{{Kimm}
  et~al.}{2019}]{Kimm:2019}
{Kimm} T.,  {Blaizot} J.,  {Garel} T.,  {Michel-Dansac} L.,  {Katz} H.,
  {Rosdahl} J.,  {Verhamme} A.,   {Haehnelt} M.,  2019, \mn@doi [\mnras]
  {10.1093/mnras/stz989}, \href
  {https://ui.adsabs.harvard.edu/abs/2019MNRAS.486.2215K} {486, 2215}

\bibitem[\protect\citeauthoryear{{Ma}, {Kasen}, {Hopkins},
  {Faucher-Gigu{\`e}re}, {Quataert}, {Kere{\v{s}}}  \& {Murray}}{{Ma}
  et~al.}{2015}]{Ma2015}
{Ma} X.,  {Kasen} D.,  {Hopkins} P.~F.,  {Faucher-Gigu{\`e}re} C.-A.,
  {Quataert} E.,  {Kere{\v{s}}} D.,   {Murray} N.,  2015, \mn@doi [\mnras]
  {10.1093/mnras/stv1679}, \href
  {https://ui.adsabs.harvard.edu/abs/2015MNRAS.453..960M} {453, 960}

\bibitem[\protect\citeauthoryear{{Nestor}, {Shapley}, {Kornei}, {Steidel}  \&
  {Siana}}{{Nestor} et~al.}{2013}]{Nestor2013}
{Nestor} D.~B.,  {Shapley} A.~E.,  {Kornei} K.~A.,  {Steidel} C.~C.,   {Siana}
  B.,  2013, \mn@doi [\apj] {10.1088/0004-637X/765/1/47}, \href
  {https://ui.adsabs.harvard.edu/abs/2013ApJ...765...47N} {765, 47}

\bibitem[\protect\citeauthoryear{Oesch et~al.,}{Oesch et~al.}{2016}]{Oesch2016}
Oesch P.~A.,  et~al., 2016, \mn@doi [The Astrophysical Journal]
  {10.3847/0004-637x/819/2/129}, 819, 129

\bibitem[\protect\citeauthoryear{{Ouchi} et~al.,}{{Ouchi}
  et~al.}{2009}]{Ouchi:2009}
{Ouchi} M.,  et~al., 2009, \mn@doi [\apj] {10.1088/0004-637X/706/2/1136}, \href
  {https://ui.adsabs.harvard.edu/abs/2009ApJ...706.1136O} {706, 1136}

\bibitem[\protect\citeauthoryear{{Pei}}{{Pei}}{1992}]{Pei1992}
{Pei} Y.~C.,  1992, \mn@doi [\apj] {10.1086/171637}, \href
  {https://ui.adsabs.harvard.edu/abs/1992ApJ...395..130P} {395, 130}

\bibitem[\protect\citeauthoryear{{Razoumov} \& {Sommer-Larsen}}{{Razoumov} \&
  {Sommer-Larsen}}{2010}]{Razoumov2010}
{Razoumov} A.~O.,  {Sommer-Larsen} J.,  2010, \mn@doi [\apj]
  {10.1088/0004-637X/710/2/1239}, \href
  {https://ui.adsabs.harvard.edu/abs/2010ApJ...710.1239R} {710, 1239}

\bibitem[\protect\citeauthoryear{{Ricotti}}{{Ricotti}}{2002}]{Ricotti:2002}
{Ricotti} M.,  2002, \mn@doi [\mnras] {10.1046/j.1365-8711.2002.05990.x}, \href
  {http://adsabs.harvard.edu/abs/2002MNRAS.336L..33R} {336, L33}

\bibitem[\protect\citeauthoryear{{Ricotti}}{{Ricotti}}{2016}]{Ricotti:2016}
{Ricotti} M.,  2016, \mn@doi [\mnras] {10.1093/mnras/stw1672}, \href
  {http://adsabs.harvard.edu/abs/2016MNRAS.462..601R} {462, 601}

\bibitem[\protect\citeauthoryear{{Ricotti} \& {Shull}}{{Ricotti} \&
  {Shull}}{2000}]{Ricotti:2000}
{Ricotti} M.,  {Shull} J.~M.,  2000, \mn@doi [\apj] {10.1086/317025}, \href
  {http://adsabs.harvard.edu/abs/2000ApJ...542..548R} {542, 548}

\bibitem[\protect\citeauthoryear{{Robertson}, {Ellis}, {Furlanetto}  \&
  {Dunlop}}{{Robertson} et~al.}{2015}]{Robertson:2015}
{Robertson} B.~E.,  {Ellis} R.~S.,  {Furlanetto} S.~R.,   {Dunlop} J.~S.,
  2015, \mn@doi [\apjl] {10.1088/2041-8205/802/2/L19}, \href
  {https://ui.adsabs.harvard.edu/abs/2015ApJ...802L..19R} {802, L19}

\bibitem[\protect\citeauthoryear{{Rosdahl}, {Blaizot}, {Aubert}, {Stranex}  \&
  {Teyssier}}{{Rosdahl} et~al.}{2013}]{Rosdahl:2013}
{Rosdahl} J.,  {Blaizot} J.,  {Aubert} D.,  {Stranex} T.,   {Teyssier} R.,
  2013, \mn@doi [\mnras] {10.1093/mnras/stt1722}, \href
  {http://adsabs.harvard.edu/abs/2013MNRAS.436.2188R} {436, 2188}

\bibitem[\protect\citeauthoryear{{Rosdahl} et~al.,}{{Rosdahl}
  et~al.}{2018}]{Rosdahl:2018}
{Rosdahl} J.,  et~al., 2018, \mn@doi [\mnras] {10.1093/mnras/sty1655}, \href
  {https://ui.adsabs.harvard.edu/abs/2018MNRAS.479..994R} {479, 994}

\bibitem[\protect\citeauthoryear{{Rosolowsky}}{{Rosolowsky}}{2005}]{Rosolowsky:2005}
{Rosolowsky} E.,  2005, \mn@doi [Publications of the Astronomical Society of
  the Pacific] {10.1086/497582}, \href
  {https://ui.adsabs.harvard.edu/abs/2005PASP..117.1403R} {117, 1403}

\bibitem[\protect\citeauthoryear{{Schaerer}}{{Schaerer}}{2002}]{Schaerer:2002}
{Schaerer} D.,  2002, \mn@doi [\aap] {10.1051/0004-6361:20011619}, \href
  {http://adsabs.harvard.edu/abs/2002A%26A...382...28S} {382, 28}

\bibitem[\protect\citeauthoryear{{Schaerer} \& {Charbonnel}}{{Schaerer} \&
  {Charbonnel}}{2011}]{Schaerer:2011}
{Schaerer} D.,  {Charbonnel} C.,  2011, \mn@doi [\mnras]
  {10.1111/j.1365-2966.2011.18304.x}, \href
  {https://ui.adsabs.harvard.edu/abs/2011MNRAS.413.2297S} {413, 2297}

\bibitem[\protect\citeauthoryear{{Schaller}, {Schaerer}, {Meynet}  \&
  {Maeder}}{{Schaller} et~al.}{1992}]{Schaller:1992}
{Schaller} G.,  {Schaerer} D.,  {Meynet} G.,   {Maeder} A.,  1992, \aaps, \href
  {https://ui.adsabs.harvard.edu/abs/1992A&AS...96..269S} {96, 269}

\bibitem[\protect\citeauthoryear{{Shapley}, {Steidel}, {Strom},
  {Bogosavljevi{\'c}}, {Reddy}, {Siana}, {Mostardi}  \& {Rudie}}{{Shapley}
  et~al.}{2016}]{Shapley:2016}
{Shapley} A.~E.,  {Steidel} C.~C.,  {Strom} A.~L.,  {Bogosavljevi{\'c}} M.,
  {Reddy} N.~A.,  {Siana} B.,  {Mostardi} R.~E.,   {Rudie} G.~C.,  2016,
  \mn@doi [\apjl] {10.3847/2041-8205/826/2/L24}, \href
  {https://ui.adsabs.harvard.edu/abs/2016ApJ...826L..24S} {826, L24}

\bibitem[\protect\citeauthoryear{{Sharma}, {Theuns}, {Frenk}, {Bower}, {Crain},
  {Schaller}  \& {Schaye}}{{Sharma} et~al.}{2016}]{Sharma2016}
{Sharma} M.,  {Theuns} T.,  {Frenk} C.,  {Bower} R.,  {Crain} R.,  {Schaller}
  M.,   {Schaye} J.,  2016, \mn@doi [\mnras] {10.1093/mnrasl/slw021}, \href
  {https://ui.adsabs.harvard.edu/abs/2016MNRAS.458L..94S} {458, L94}

\bibitem[\protect\citeauthoryear{{Teyssier}}{{Teyssier}}{2002}]{Teyssier:2002}
{Teyssier} R.,  2002, \mn@doi [\aap] {10.1051/0004-6361:20011817}, \href
  {http://adsabs.harvard.edu/abs/2002A%26A...385..337T} {385, 337}

\bibitem[\protect\citeauthoryear{{Vacca}, {Garmany}  \& {Shull}}{{Vacca}
  et~al.}{1996}]{Vacca:1996}
{Vacca} W.~D.,  {Garmany} C.~D.,   {Shull} J.~M.,  1996, \mn@doi [\apj]
  {10.1086/177020}, \href {http://adsabs.harvard.edu/abs/1996ApJ...460..914V}
  {460, 914}

\bibitem[\protect\citeauthoryear{{Vanzella} et~al.,}{{Vanzella}
  et~al.}{2012}]{Vanzella2012}
{Vanzella} E.,  et~al., 2012, \mn@doi [\apj] {10.1088/0004-637X/751/1/70},
  \href {https://ui.adsabs.harvard.edu/abs/2012ApJ...751...70V} {751, 70}

\bibitem[\protect\citeauthoryear{{Vanzella} et~al.,}{{Vanzella}
  et~al.}{2016}]{Vanzella:2016}
{Vanzella} E.,  et~al., 2016, \mn@doi [\apj] {10.3847/0004-637X/825/1/41},
  \href {https://ui.adsabs.harvard.edu/abs/2016ApJ...825...41V} {825, 41}

\bibitem[\protect\citeauthoryear{{Vanzella} et~al.,}{{Vanzella}
  et~al.}{2018}]{Vanzella:2018}
{Vanzella} E.,  et~al., 2018, \mn@doi [\mnras] {10.1093/mnrasl/sly023}, \href
  {https://ui.adsabs.harvard.edu/abs/2018MNRAS.476L..15V} {476, L15}

\bibitem[\protect\citeauthoryear{{Weingartner} \& {Draine}}{{Weingartner} \&
  {Draine}}{2001}]{Weingartner:2001}
{Weingartner} J.~C.,  {Draine} B.~T.,  2001, \mn@doi [\apj] {10.1086/318651},
  \href {https://ui.adsabs.harvard.edu/abs/2001ApJ...548..296W} {548, 296}

\bibitem[\protect\citeauthoryear{{Wise} \& {Cen}}{{Wise} \&
  {Cen}}{2009}]{Wise:2009}
{Wise} J.~H.,  {Cen} R.,  2009, \mn@doi [\apj] {10.1088/0004-637X/693/1/984},
  \href {https://ui.adsabs.harvard.edu/abs/2009ApJ...693..984W} {693, 984}

\bibitem[\protect\citeauthoryear{{Wise}, {Demchenko}, {Halicek}, {Norman},
  {Turk}, {Abel}  \& {Smith}}{{Wise} et~al.}{2014}]{Wise:2014}
{Wise} J.~H.,  {Demchenko} V.~G.,  {Halicek} M.~T.,  {Norman} M.~L.,  {Turk}
  M.~J.,  {Abel} T.,   {Smith} B.~D.,  2014, \mn@doi [\mnras]
  {10.1093/mnras/stu979}, \href
  {http://adsabs.harvard.edu/abs/2014MNRAS.442.2560W} {442, 2560}

\bibitem[\protect\citeauthoryear{{Xu}, {Wise}, {Norman}, {Ahn}  \&
  {O'Shea}}{{Xu} et~al.}{2016}]{Xu:2016}
{Xu} H.,  {Wise} J.~H.,  {Norman} M.~L.,  {Ahn} K.,   {O'Shea} B.~W.,  2016,
  \mn@doi [\apj] {10.3847/1538-4357/833/1/84}, \href
  {https://ui.adsabs.harvard.edu/abs/2016ApJ...833...84X} {833, 84}

\bibitem[\protect\citeauthoryear{{Yajima}, {Choi}  \& {Nagamine}}{{Yajima}
  et~al.}{2011}]{Yajima:2011}
{Yajima} H.,  {Choi} J.-H.,   {Nagamine} K.,  2011, \mn@doi [\mnras]
  {10.1111/j.1365-2966.2010.17920.x}, \href
  {https://ui.adsabs.harvard.edu/abs/2011MNRAS.412..411Y} {412, 411}

\makeatother
\end{thebibliography}

\bsp	
\label{lastpage}
\end{document}